\documentclass{aa}

\usepackage{amsmath, amssymb, amsbsy, amsfonts}
\usepackage{txfonts, placeins, hyperref, breakurl}
\usepackage{rotating, array, supertabular, graphicx, natbib}
\usepackage{paralist}
\usepackage{booktabs,xspace}
\usepackage{multirow}
\usepackage[table]{xcolor}
\usepackage{rotating}
\usepackage{lscape}

\renewcommand{\vec}[1]{\ensuremath{\mathchoice{\mbox{\boldmath$\displaystyle#1$}}
{\mbox{\boldmath$\textstyle#1$}}
{\mbox{\boldmath$\scriptstyle#1$}}
{\mbox{\boldmath$\scriptscriptstyle#1$}}}}

\newcommand{\ea}[1]{{\it et al.}}
\newcommand{\sv}[1]{$\sigma_{\rm var}$}

\begin{document}

\title{Gaia reference frame amid quasar variability and proper motion patterns in the data}

\author{R. K. Bachchan\inst{1}, D. Hobbs\inst{1}\and L. Lindegren\inst{1}}
\institute{Lund Observatory, Department of Astronomy and Theoretical Physics, Lund University, Box 43, 22100 Lund, Sweden\\
            \email{rajesh, david, lennart@astro.lu.se}}
\date{Received 9 December 2015 / Accepted x March 2016}

\abstract
{Gaia's very accurate astrometric measurements will allow the optical realisation of the International Celestial Reference System to be improved by a few orders of magnitude. Several sets of quasars are used to define a kinematically stable non-rotating reference frame with the barycentre of the solar system as its origin. Gaia will also observe a large number of galaxies. Although they are not point-like, it may be possible to determine accurate positions and proper motions for some of their compact bright features.}
{The optical stability of the quasars is critical, and we investigate how accurately the reference frame can be recovered. Various proper motion patterns are also present in the data, the best known is caused by the acceleration of the solar system barycentre, presumably, towards the Galactic centre. We review some other less well-known effects that are not part of standard astrometric models.}
{We modelled quasars and galaxies using realistic sky distributions, magnitudes, and redshifts. Position variability was introduced using a Markov chain model. The reference frame was determined using the algorithm developed for the Gaia mission, which also determines the acceleration of the solar system. We also tested a method for measuring the velocity of the solar system barycentre in a cosmological frame.}
{We simulated the recovery of the reference frame and the acceleration of the solar system and conclude that they are not significantly disturbed by quasar variability, which is statistically averaged. However, the effect of a non-uniform sky distribution of the quasars can result in a correlation between the parameters describing the spin components of the reference frame and the acceleration components, which degrades the solution. Our results suggest that an attempt should be made to astrometrically determine the redshift-dependent apparent drift of galaxies that is due to our velocity relative to the cosmic microwave background, which in principle could allow determining the Hubble parameter.}
{}

\authorrunning{R. K. Bachchan}
\titlerunning{The Gaia reference frame amid quasar variability and proper motion patterns in the data}
\maketitle

\section{Introduction}\label{Sec:intro}
Gaia is an astrometric satellite launched in late 2013 and designed to produce a three-dimensional map of the local part of our Galaxy from which our Galaxy's composition, formation, and evolution can be reconstructed. The satellite will measure the positions, proper motions, and parallaxes of at least one billion stars in the Milky Way. In addition, Gaia will detect thousands of exoplanets, asteroids, and about half a million distant quasars in the optical spectrum. Since the accuracy is at the micro-arcsecond level, these precise measurements help to improve the optical realisation of the International Celestial Reference System (ICRS) by a few orders of magnitude compared to its current realisation by the Hipparcos and Tycho catalogues \citep{1997ESASP1200.....E}. It will also provide a number of new tests of the general theory of relativity. 

The ICRF \citep{1998AJ....116..516M} is a quasi-inertial reference frame that was originally defined by the measured positions of 212 extragalactic radio sources derived from ground-based very long baseline interferometry (VLBI) and has its reference point at the barycentre of the solar system. In general relativity there is no true inertial reference frame; the extragalactic sources (quasars) used to define the ICRF are so far away, however, that any net angular motion is almost zero. The ICRF is now the standard reference frame used to define the positions of astronomical objects. It has been adopted by International Astronomical Union in 1998 \citep{1998AJ....116..516M}. In 2009, the second realisation, ICRF2 \citep{2009ITN....35....1M, 2015AJ....150...58F}, was adopted. This included improved models and concepts and was based on  3414 compact astronomical sources. ICRF2 defines the reference frame to an accuracy of 40 $\mu$as and includes 295 defining sources uniformly distributed on the sky and selected on the basis of positional stability and the lack of extensive intrinsic source structure.

The Hipparcos and Tycho catalogues currently serve as the corresponding optical realisation of the International Reference System (ICRS), but they will be superseded by the Gaia mission in the coming years \citep{Mignard2011}. This frame must be constructed with the same principles as the ICRS, that is, overall a kinematically non-rotating system with the same orientation as the radio ICRF. It is estimated that Gaia will astrometrically measure some 500\,000 quasars, and their repeated measurement over the estimated five year mission will lead to a new kinematically defined inertial reference frame in the optical. 

Quasars located at the centre of distant active galaxies are characterised by extremely compact and bright emission. They are at cosmological distances and therefore show negligible transverse motion. However, recent observations \citep{2011A&A...526A..25T,2009A&A...505L...1P,2008A&A...483..759K} of active galactic nuclei (AGNs) and theoretical studies \citep{2012A&A...538A.107P} indicate that variability in the accretion disk and dusty torus surrounding the central black hole can cause photocentre shifts of up to the milliarcsec level, therefore it is probable that quasar variability will affect the reference frame as well. This paper investigates the statistical impact of variability on the Gaia reference frame based on simulated astrometric observations.

Another interesting aspect is the acceleration of the solar system
that was first pointed out by \cite{1995ESASP.379...99B}. It causes a pattern of proper motions that can be solved for while determining the reference frame orientation and spin parameters. We present an estimate of how well the acceleration vector can be determined based on realistic Gaia simulations that also assess the impact of quasar variability. In addition to this effect, there may be several more subtle effects present in the real data. Many of these will be weak; nevertheless, it is interesting to review and compare them, and if they are deemed strong enough, we can simulate them together with the acceleration of the solar system to assess their impact on the solution and the potential to distinguish the different effects.

The organisation of the paper is as follows: Sect.~\ref{Sec:perspective} presents various proper motion effects that could have an impact on the Gaia measurements, and we quantify their respective magnitudes. In Sect.~\ref{Sec:sim} we present the numerical simulations. In Sect.~\ref{Sec:refframe} we explain how the Gaia reference frame is determined together with the patterns of proper motion discussed in Sect.~\ref{Sec:perspective}. In Sect.~\ref{Sec:results} the simulation results are presented and discussed, and finally in Sect.~\ref{Sec:conclusions} conclusions are given.

\begin{table}[ht]
\centering
\caption[]{Table of various perspective effects. o refers to an effect that is due to the observer's motion. s refers to an effect that is due to the source's motion, and c refers to an effect that is of cosmological origin. We arbitrarily define the boundary between low and high $z$ as a redshift of 1.0. \label{Tab:effects}}
\rotatebox{90}{
\begin{tabular}{llllll}
	\toprule\toprule
	Effect &
	Origin &
	Description &
	Measurement method&
	Expected magnitude&
	Redshift dependence \\
	\midrule[0.2pt]
	Acceleration of             & o     & Acceleration of the solar system assumed to be towards       & Astrometric          & $\sim$ 4.3~$\mu$as~yr$^{-1}$          & None \\
	the solar system            &       & the Galactic centre resulting in patterns of proper motion.  &                      &                                       & \\
	                            &       & However, the local group of galaxies and clusters of local   &                      &                                       & \\
	                            &       & supercluster will also contribute to the effect.             &                      &                                       & \\
	\midrule[0.2pt]
	Cosmological                & o     & Instantaneous velocity of the solar system with respect      & CMB and astrometric  & 1--2~$\mu$as~yr$^{-1}$                & Increases at low $z$ \\ 
	proper motion               &       & to the CMB can cause distant extragalactic sources to        &                      &                                       & \\
	                            &       & undergo an apparent systematic proper motion.                &                      &                                       & \\
	\midrule[0.2pt]
	Gravitational waves         & o     & Primordial gravitational waves produce systematic            & Dedicated detectors/ & Unknown but                           & None \\
	                            &       & proper motions over the sky.                                 & possibly astrometric & probably $<$ 1~$\mu$as~yr$^{-1}$      & \\
	\midrule[0.2pt]
	Transverse redshift drift/  & c     & A temporal shift of the angular separation of distant        & Astrometric          & 0.2~~$\mu$as~yr$^{-1}~$(Bianchi)      & Increases at high $z$, \\ 
	cosmic parallax             &       & sources can be used to detect an anisotropic expansion       &                      & 0.02~$\mu$as~yr$^{-1}~$(LTB)          & decreases at very high $z$\\
	                            &       & of the Universe and results in a pattern of proper motions.  &                      &                                       & \\
	\midrule[0.2pt]
	Peculiar proper motion      & s     & Proper acceleration is the observed transverse acceleration  & Astrometric          & Can be $10$~$\mu$as~yr$^{-1}$         & None \\
	                            &       & of an object due to the local gravitational field.           &                      & for Galactic clusters                 & \\
	\midrule[0.2pt]
	Quasar microlensing         & s     & Weak microlensing can induce apparent motions of quasars.    & Astrometry           & 10's of~$\mu$as~yr$^{-1}$ but         & None \\
	                            &       &                                                              & photometry           & is extremely rare                     & \\
	\bottomrule
\end{tabular}
}
\end{table}

\section{Proper motion effects in the Gaia data}\label{Sec:perspective}
\cite{2010ivs..conf...60T} noted that individual apparent proper motions of distant radio sources are generally attributed to the internal structure of AGNs. However, he pointed out that there are a number of other effects that could give rise to systematic apparent motions of quasars, and we briefly consider these here. The various sources of proper motion are summarised in Table \ref{Tab:effects}. We note that the terminology used in the literature describing the different effects varies and can be inconsistent,
which is mainly due to the historical progression of the topics.  
\subsection{Photocentre variability}\label{photo_var}
\cite{2014ApJ...789..166P} considered the various sources of quasar instability. They concluded that the most important effect is optical photo-centric motion, where the internal structure of the AGN could result in variability of typically less that 60~$\mu$as but up to 100~$\mu$as in extreme cases \citep{2011A&A...526A..25T}. In the present simulations we considered the extreme case where these effects combine to produce maximum variability distortions of 100~$\mu$as. The timescales of quasar variability have been studied by \cite{1993AJ....105..437S} and \cite{2011A&A...526A..25T} and were found to range from 3--15 years, peaking between 6--9 years. We chose to use 2 and 10 years, which roughly represents the range of timescales to which Gaia is most sensitive, assuming a mission duration of 5 years.

\clearpage
\subsection{Acceleration of the solar system} 
The solar system orbital velocity of $\sim$200~km~s$^{-1}$ around the Galactic centre results in an aberration effect of about 2.5 arcminutes \citep{2014ApJ...789..166P} in the direction of motion. The acceleration of the solar system in its Galactic orbit causes this effect to change slowly, which results in a proper motion pattern for all objects on the sky \citep{1995ESASP.379...99B, 2003A&A...404..743K, 1997ApJ...485...87G, 1998RvMP...70.1393S, 2003AJ....125.1580K, 2006AJ....131.1471K, 2014MNRAS.445..845M}. This effect has also been referred to as secular aberration drift \citep{2010ivs..conf...60T}. The effect is generally assumed to be towards the Galactic centre where most of the mass is concentrated, but of course the unknown distribution of dark matter may affect the direction. The Galactocentric acceleration can be calculated as $a = v^2/r$. For example, assuming a circular velocity $v = 223$~km~s$^{-1}$ and radius $r = 8.2$~kpc \citep{2008ApJ...689.1044G}, the resulting proper motion pattern is
\begin{equation}\label{a_tilde}
\mu = \tilde{a}\sin\theta\ , \quad \tilde{a} = \frac{a}{c} = 4.27~\text{$\mu$as~yr$^{-1}$}\ ,
\end{equation}
where $c$ is the speed of light and $\theta$ is the angular distance between the object and the Galactic centre \citep{1995ESASP.379...99B}. 
\begin{table}[h]
        \centering \caption[]{Values of proper motion, $\tilde{a}$, that are due to the extragalactic acceleration of various local group galaxies and some clusters of the local supercluster. \label{Tab:appA}} 
        \begin{tabular}{l*{2}{c}c}
                \toprule
                Galaxy          &  $M$                    & $d$    &$\tilde{a}$                 \\
                &  ($M_\odot$)  & (kpc)                   &($\mu$as~yr$^{-1}$)                  \\
                \midrule[0.2pt]
                LMC             &  2.0$\times10^{10}$     & 50     &     2.4$\times10^{-2}$      \\
                M31             &  4.0$\times10^{11}$     & 760    &     2.1$\times10^{-3}$      \\
                Sagittarius     &  1.5$\times10^{08}$     & 24     &     7.6$\times10^{-4}$      \\
                SMC             &  8.0$\times10^{08}$     & 59     &     7.0$\times10^{-4}$      \\
                M33             &  1.4$\times10^{10}$     & 830    &     6.1$\times10^{-5}$      \\
                NGC6822         &  1.9$\times10^{09}$     & 500    &     2.3$\times10^{-5}$      \\
                UrsaMinor       &  1.7$\times10^{07}$     & 63     &     1.3$\times10^{-5}$      \\
                \midrule[0.2pt]
                Cluster         &                         &        &                    \\
                \midrule[0.2pt]
                Virgo           &  1.2$\times 10^{15}$    & 16036  & 1.4$\times10^{-2}$ \\
                Fornax          &  4.0$\times 10^{13}$    & 20491  & 2.9$\times10^{-4}$ \\
                Norma           &  1.0$\times 10^{15}$    & 69937  & 6.2$\times10^{-4}$ \\
                Antlia          &  3.3$\times 10^{14}$    & 41561  & 5.8$\times10^{-4}$ \\
                Coma            &  7.0$\times 10^{14}$    & 102900 & 2.0$\times10^{-4}$ \\
                \bottomrule
        \end{tabular}
        \tablefoot{The masses and the distances of galaxies are taken from \cite{VandenBergh:991671}.
                The distance of clusters are calculated from redshift data available at NED (Nasa Extragalactic Database: \url{http://ned.ipac.caltech.edu/}) and the masses are taken from \citep{2001A&A...375..770F, 2011A&A...532A.104N, 1996ApJ...467..168B, 1985A&AS...61...93H, 2009A&A...498L..33G}.}
\end{table}

The local group and local supercluster also accelerate the barycentre of the solar system similar to the Galactocentric acceleration. We estimated $\tilde{a}$ for some 33 galaxies of the local group, using $a = GM/d^2$ . The seven largest examples are listed in Table~\ref{Tab:appA}. The values of $\tilde{a}$ are very low ($< 0.1~\mu$as~yr$^{-1}$), and Gaia will not be able to measure
them. We similarly calculated the value of $\tilde{a}$ for some clusters of the local supercluster, and in this case the values were found to be very low as well and not measurable by Gaia.

If we consider some non-axisymmetric potential for the Milky Way that is due to the bar and the spiral arms, then the acceleration vector will not point exactly towards the Galactic centre. For example, in the potential used by \citet{2013ApJ...768..152F}, we find the acceleration vector to be offset by about $2^{\circ}$.

\subsection{Cosmological proper motion}\label{cos_prop_mot}
The instantaneous velocity of the solar system with respect to the cosmological microwave background (CMB) will cause extragalactic sources to undergo an apparent systematic proper motion. The effect is referred to as cosmological or parallactic proper motion \citep{1986AZh....63..845K}. This effect depends on the redshift $z,$ and fundamental cosmological parameters can in principle be determined from the motion. The velocity ($v$) of the solar system with respect to the observable Universe produces a dipole pattern in the CMB temperature with $\Delta T/T = v/c$. Observations of the CMB indicate that $v = 369 \pm 0.9$ km~s$^{-1}$ in the apex direction with Galactic longitude $l = 263.99^{\circ}$ and latitude $b = 48.26^{\circ}$ \citep{2009ApJS..180..225H, 2014A&A...571A..27P}. This motion should produce a parallactic shift of all extragalactic objects towards the antapex at the angular rate
\begin{equation}\label{mu_cosm}
\mu = \frac{v}{d} \sin\beta\ , 
\end{equation}
where $\beta$ is the angle between the source and apex directions, and 
\begin{equation}\label{dist}
d(z) = \frac{c}{H_0}\int_0^z \frac{dz^\prime}{\sqrt{\Omega_m(1 + z)^3 + \Omega_r(1 + z)^4 + \Omega_k(1 + z)^2 + \Omega_\Lambda}}\ \end{equation}
\begin{figure}[h]\label{Fig:figdist}
        \vspace{0.0cm} \centering
        \includegraphics[height = 5.5cm, trim = {0cm 7cm 0cm 7cm}]{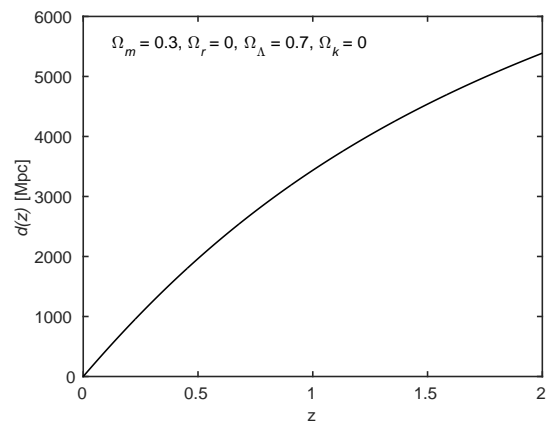}
        \caption{Variation of comoving distance with redshift.\label{Fig:dist}}
\end{figure}
is the transverse comoving distance, which for a flat universe equals the line-of-sight comoving distance, illustrated in Fig.~\ref{Fig:figdist}. \citep{1999astro.ph..5116H,2006gere.book.....H,2008cosm.book.....W}. The quantities $\Omega_m$, $\Omega_r$, $\Omega_k$, and $\Omega_\Lambda$ are the dimensionless energy densities of matter, radiation, curvature, and cosmological constant, respectively. For very low redshift ($z\ll 1$), $\mu \simeq (vH_0/cz)\sin\beta$, or 1--2~$\mu$as~yr$^{-1}$ for $z = 0.01$ \citep{1986AZh....63..845K}. In principle, this could be within the reach of Gaia, depending on the number of available objects and the precision of the observations.

Both the acceleration of the solar system and the cosmological proper motion give rise to dipole patterns in the proper motions of distant objects. However, the former does not depend on the redshift, while the latter does, which makes it possible to separate the effects.

\subsection{Primordial gravitational waves}
Primordial gravitational waves could give rise to systematic proper motions over the sky, composed of second-order transverse vector spherical harmonics \citep{1997ApJ...485...87G}. Changes in the space time metric that are due to gravitational waves alter the optical path length, preserve the sources brightness and intensity, but produce oscillations in the apparent position of the source. If the interval of observation is shorter than the period of the gravitational wave, then this will be seen as a systematic proper motion on the sky. \cite{1996ApJ...465..566P} and \cite{1997ApJ...485...87G} have investigated the low frequency observational constraints on gravitational waves
in detail, which could arise naturally in inflationary theories of cosmology. It is anticipated that Gaia is not sufficiently sensitive to detect primordial gravitational waves directly, but it should be able to place upper limits on their energy density, which is comparable to that of pulsar timing measurements \citep{2011PhRvD..83b4024B}.

\subsection{Transverse redshift drift / cosmic parallax}
\cite{1966ApJ...143..379K} and \cite{2011A&A...529A..91T} pointed out that an anisotropic expansion of the Universe would result in a distortion, as a function of redshift, of all distant objects on the celestial sphere in a particular direction. The effect gives a direct measurement of space time curvature, which is similar to a gravitational lens, but in this case is due to the cosmological curvature and not to a single body. The time-dependent components of the distortion would result in patterns of proper motion that could be a function of the redshifts and can be measured in principle. \cite{2011A&A...529A..91T} has shown that the dipole term does not vary significantly and agrees with the predicted estimates of 4--6~$\mu$as yr$^{-1}$. For the quadrupole anisotropy, which could be interpreted as an anisotropic Hubble expansion or as an indicator of primordial gravitational waves, no detection has been made.

\cite{2009PhRvD..80f3527Q, 2012PhR...521...95Q} used the term cosmic parallax for the varying angular separation between any two distant sources, caused by the anisotropic expansion of the Universe. They considered two different scenarios:
\begin{itemize}
\item Bianchi 1 models in which the observer is centrally embedded in an intrinsically anisotropic expansion of the early Universe. In this case, the anisotropic stress of dark energy can induce an anisotropic expansion of the Universe at late times that cannot be constrained by the CMB background measurements. Their simplified model, assuming an anisotropy of about 1\%, gives a proper motion pattern of about 0.2~$\mu$as~yr$^{-1}$.
\item Lemaitre-Tolman-Bondi (LTB) void models in which the observer is off-centre and the Universe is inhomogeneous and isotropic. For the models considered by \cite{2012PhR...521...95Q}, they derived effects of about 0.02~$\mu$as~yr$^{-1}$. 
\end{itemize}

\subsection{Peculiar proper motion}
In analogy to the peculiar motion of stars, which is defined as the star velocity relative to the local standard of rest, the peculiar motion of a galaxy is its velocity relative to the Hubble flow. The transverse component of this motion causes a proper motion of the galaxy that we call peculiar proper motion. The typical transverse velocity ($v_{\text{pec}}$) is of $\sim600$~km~s$^{-1}$, but in rich galaxy clusters it may be as high as $\sim1500$~km~s$^{-1}$ \citep{2000gaun.book.....S}. 
The proper motion is then given by
\begin{equation}\label{mu}
\mu_{\text{pec}} = \frac{v_{\text{pec}}}{d(z)}\ ,
\end{equation}
where $d(z)$ is the comoving distance Eq.~\eqref{dist} and $\mu_{\text{pec}}\simeq v_{\text{pec}}H_0/cz$ for low redshifts. This gives a peculiar proper motion in the range 3--7~$\mu$as~yr$^{-1}$ at $z = 0.01$. The effect is random and is not expected to give a systematic pattern of proper motion.

\subsection{Quasar microlensing} 

\cite{2002MNRAS.331..649B} pointed out that microlensing events can be detected by Gaia both from the brightening of the source star (photometric microlensing) and the positional excursion (astrometric microlensing). The optical depth is given by them as $10^{-7}$ and $2.5 \times 10^{-5}$ for photometric and astrometric microlensing, respectively, which corresponds to about $1300$ photometric and $25000$ astrometric events during the course of the nominal five-year Gaia mission. The individual images of microlensed sources cannot be resolved by Gaia, but the centroid shift along the trajectory of a source can be measured. In the case of stellar microlensing of quasars, the centroid shift can be tens of $\mu$as \citep{1998ApJ...501..478L}.

\citet{2011ARep...55..954S} derived detailed expressions for the apparent proper motions caused by weak microlensing. In Appendix \ref{Sec:appB} we show that the statistical effect is related to the optical depth for photometric microlensing. If we consider that a source is a quasar with zero proper motion ($\mu_Q = 0$), then by considering a large number of lensing objects with random proper motions, we find that
\begin{equation}\label{muRMS}
\left(\mu_{Q'} \right)_{\text{RMS}} = \left(\mu_{L} \right)_{\text{RMS}}\times \sqrt{\tau} \, ,
\end{equation}
where $\left(\mu_{Q'} \right)_{\text{RMS}}$ is the RMS value of the quasar proper motion, $\left(\mu_{L} \right)_{\text{RMS}}$ is the RMS value of the lens proper motion, and $\tau$ is the probability that the source $Q$ is within the Einstein radius of some lens, which is simply the  optical depth for photometric microlensing.

If we assume that the number of galaxies is $0.05$ Mpc$^{-3}$, the lens mass is $10^{11}$ M$_\odot$, the lens distance is 5 Gpc, and the source distance is 10 Gpc, then the value of $\tau \sim0.08$ in the extragalactic case. Similarly, in case of Galactic microlensing, if we take the number of sources as $0.1$ pc$^{-3}$, the mass as $1$ M$_\odot$, the lens distance as $5$ kpc, and the source distance as 10 kpc, then $\tau \sim 10^{-6}$. Clearly, $\left(\mu_{Q'} \right)_{\text{RMS}}\ll1~\mu$as~yr$^{-1}$ in both cases. This means that even though a distant galaxy acting as a lens may have a tiny motion in the transverse direction with respect to the observer, the source quasar being at rest, the apparent proper motion of the quasar is too weak to be measured by Gaia, and the same applies to Galactic microlensing.

\subsection{Other apparent motions of quasars} 
The peculiar proper motion of quasars is expected to be very small because of their large distances. For example, for a quasar at $z = 3$, $\mu_{\text{pec}} = 0.02~{\rm \mu as~yr^{-1}}$ according to Eqs.~\eqref{dist} and \eqref{mu}. This proper motion is very small and hence the quasars can in this respect be assumed to be stationary. However, radio observations of quasars have shown that a significant number of them have apparent proper motions exceeding $50~{\rm \mu as~yr^{-1}}$ \citep{2005ASPC..340..477M, 2011A&A...529A..91T}. \cite{2011ARep...55..954S} have pointed out several possible causes, including apparent superluminal motions in radio jets, gravitational waves and weak microlensing by stars and dark bodies in our Galaxy. \cite{2011ARep...55..954S} has given a number of examples of apparent motions that are due to microlensing, and the order of proper motions in some cases has been shown to be several tens of~$\mu$as~yr$^{-1}$. However, the number of such events is estimated to be very small. 

\section{Simulations} \label{Sec:sim}
\subsection{Data sets} \label{Sec:sim:datasets}
To simulate the quasars and galaxies, we took the positions, magnitudes, and redshifts of the objects from the GUMS \citep{2012A&A...543A.100R} simulated data set. This data set includes a realistic sky, magnitude, and redshift distribution of about 1 million quasars and 38 million unresolved galaxies\footnote{`Unresolved' here means that individual stars in a galaxy cannot be seen by Gaia. The galaxy as a whole is however often an extended object at the resolution of Gaia.} expected to be seen by Gaia. As a significant fraction of the galaxies will not be observed by Gaia because of their extended structure, we focus on the simpler and most point-like objects down to $G = 20$ magnitude by selecting only ellipticals and spirals (Hubble types E2, E-S0, Sa, and Sb with a bulge-to-total flux ratio of 1.0, 0.9, 0.56--0.57, and 0.31--0.32, respectively) in the redshift range 0.001 to 0.03. This results in just over 100\,000 objects, which is sufficient for our simulations. 

The GUMS data set of nearly 1 million quasars represents an idealised case of\ those objects that in principle could be used for the frame determination assuming that they can be correctly classified by means of the Gaia observations \citep{2008MNRAS.391.1838B}. We randomly selected half a million quasars from the full dataset for our simulations. 

As a very conservative alternative, we also considered using the Initial Gaia Quasar List \citep[IGQL;][]{2009A&A...505..385A, 2012sf2a.conf...61A}, which currently consists of about 1.2 million quasars compiled from various catalogues. It will be used in the early Gaia data processing to simplify the identification of quasars. IGQL contains a snapshot of the best information of optical quasars available just before the launch of Gaia. The quasars in the IGQL are not uniformly distributed on the sky, but there are a number of bands and high-density regions corresponding to the various surveys used to compile it \citep{2010A&A...518A..10V, 2012A&A...537A..99S, 2009A&A...505..385A, 2012sf2a.conf...61A, 2011ApJS..194...45S, 2014A&A...563A..54P}. Notably, the IGQL is significantly lacking sky coverage in the southern hemisphere, and the impact of this non-uniform distribution is discussed in Sect.~\ref{Sec:results}.

For both quasar data sets, a random subset of about 3300 objects
were assumed to have accurate VLBI positions that simulate the
positions in the ICRF2 catalogue.

\subsection{Simulating quasar observations} \label{Sec:sim:quasar}  
To include the Galactocentric acceleration in these data sets, we simulated a pattern of proper motions 
of the form \citep{2006AJ....131.1471K}\begin{equation}\label{Eq:sim1}
\begin{aligned}
\mu_{\alpha^\star} & = -\tilde{a}_1 \sin\alpha + \tilde{a}_2 \cos\alpha \\
\mu_\delta         & = -\tilde{a}_1 \cos\alpha\sin\delta - \tilde{a}_2 \sin\alpha \sin\delta + \tilde{a}_3 \cos\delta  \, ,
\end{aligned}
\end{equation}
where $\mu_{\alpha^\star} = \mu_\alpha \cos\delta$ and $\vec{\tilde{a}} = (\tilde{a}_1, \tilde{a}_2, \tilde{a}_3)$ is the acceleration divided by the speed of light, Eq.~\eqref{a_tilde}, in ICRS. We assumed acceleration components of $\vec{\tilde{a}} = (-0.236, -3.756, -2.080)$~$\mu$as~yr$^{-1}$ in ICRS, which corresponds to $(4.3, 0, 0)$~$\mu$as~yr$^{-1}$ in Galactic coordinates, that
is,\ directed towards the Galactic centre.

\begin{figure*}[t]\label{Fig:fig2}
\vspace{0.0cm} \centering
\includegraphics[height=6.7cm, trim = {0.6cm 0.7cm 1cm 1cm}, clip]{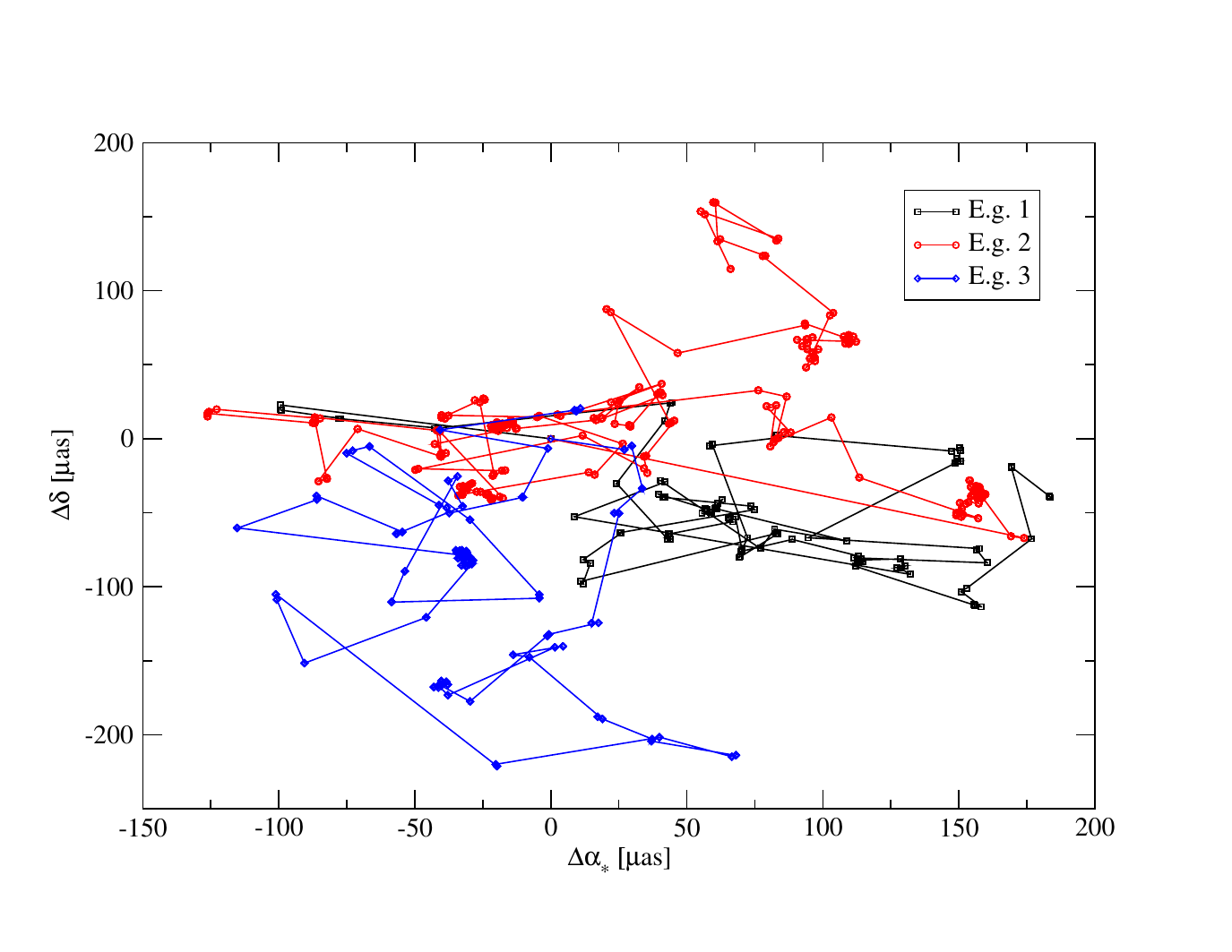}
\includegraphics[height=6.7cm, trim = {0.6cm 0.7cm 1cm 1cm}, clip]{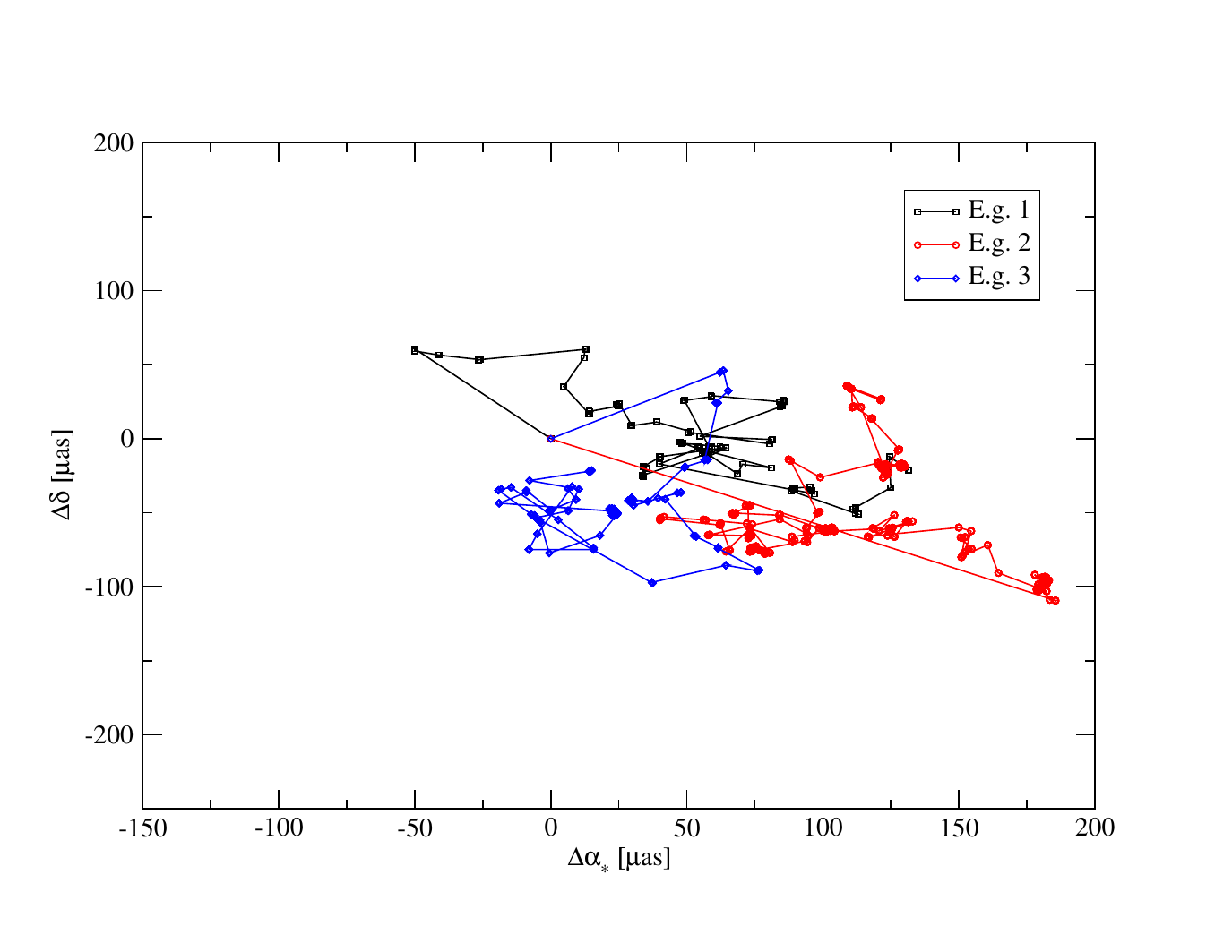}
\caption{Markov chain photo-centre variability for a VLBI quasar with timescales, $\tau_{\rm cor}$, of 2 (left) and 10 years (right). In each case three examples are given for the same sources.\label{Fig:markov}}
\end{figure*}

To simulate quasar photocentric variability, we used a Markov chain with an exponentially decaying correlation \citep{2011A&A...532A..13P, chatfield2013analysis, 1942AnMat..43..351D} with a characteristic correlation timescale $\tau_{\rm cor}$. This results in random variations in positions at time $t_i$ that are both Gaussian and Markovian and generated by
\begin{equation}\label{Eq:sim2}
\begin{bmatrix}
\Delta\alpha_*(t_i) \\ 
\Delta\delta(t_i)   \\ 
\end{bmatrix} = e^{-\Delta t_i/\tau_{\rm cor}}
\begin{bmatrix}
\Delta\alpha_*(t_{i-1}) \\ 
\Delta\delta(t_{i-1})   \\ 
\end{bmatrix} + 
\begin{bmatrix}
g_i^{\alpha_*} \\ 
g_i^{\delta}   \\ 
\end{bmatrix}\, ,
\end{equation}
where $\Delta t_i = t_i - t_{i - 1}$ and the $g_i^{\alpha_*}$ and $g_i^{\delta}$ are sampled from a Gaussian distribution with zero mean and standard deviation
\begin{equation}\label{Eq:sim3}
\sigma_i = \sigma_{\text{var}}\sqrt{1- \exp(-2\Delta t_i/\tau_{\rm cor})}\, .
\end{equation}
$\sigma_{\text{var}}$ is the standard deviation of the random variations $\Delta\alpha_*(t)$ and $\Delta\delta(t)$, which are used to perturb the source positions for each quasar independently. We assumed photocentre variations of $\sigma_{\text{var}} = 100~\mu$as and $\tau_{\rm cor} = 2$ and $10$~years (Sect.~\ref{photo_var}). Two examples of Markov chain photo-centre variability are shown in Fig.~\ref{Fig:markov}.

We did not simulate the offset of the optical positions of VLBI quasars from their radio positions. This could amount to tens of milliarcseconds or more \citep{2012MmSAI..83..952M, 2013A&A...553A..13O, 2014AJ....147...95Z}, which could strongly
affect the determination of the orientation of the optical reference frame. However, since we are primarily interested in determining proper motion patterns, only the time variation of the offset matters. We assumed that this variation is much smaller and covered by the Markov chain model. 

\subsection{Simulating galaxy observations} \label{Sec:sim:galaxies}
\begin{figure*}[t]
\vspace{0.0cm} \centering
\includegraphics[height = 12cm, trim = {1cm 1cm 0cm 1cm}, clip]{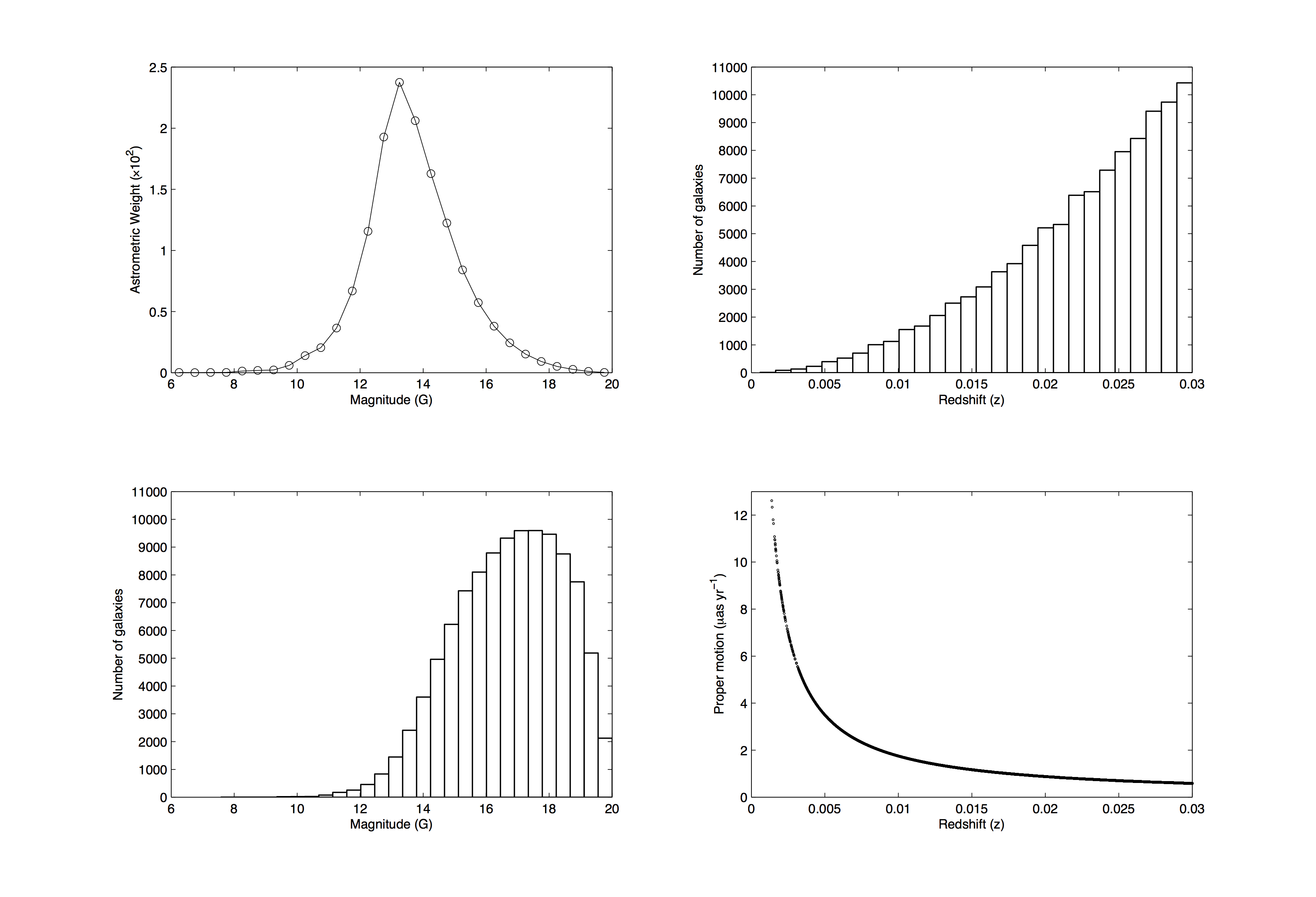}
\caption{Left: the astrometric weight ($w$) of GUMS galaxy observations \citep[see][]{IAU:6911396}, where $w=\sum_{\rm obs}\sigma_{\rm obs}^{-2}$,
is the total statistical weight of the assumed along-scan standard error of Gaia observations. Below is the number of GUMS galaxies as a function 
of $G$ magnitude. The two left plots show that the highest astrometric weight is from $G=12$ to $G=16,$ but the number of galaxies 
peaks around $G=17$. 
Right: the redshift distribution of the galaxies from GUMS and the magnitude of the cosmological proper motion effect as a function of redshift.\label{Fig:distal}}
\label{Fig:weightGalaxies}
\end{figure*}

As explained in Sect.~\ref{Sec:sim:datasets}, we simulated only 100\,000 galaxies ({\tiny GAIA\_GALAXY}) from GUMS, which were used for simulations involving redshift-dependent proper motion patterns. This number is very conservative considering there are 38 million galaxies in the GUMS catalogue, but we also restricted ourselves to a narrow redshift range (0.001 -- 0.03) where the effect is strongest (see Fig. \ref{Fig:weightGalaxies}). In this redshift range the angular diameter of a typical galaxy core (assuming a core diameter of~1~kpc) is between $46\arcsec$ and $1.5\arcsec$, respectively. In the simulation we assumed that these galaxies can be observed with the same accuracies as point-like sources of the same total magnitude. This assumption is certainly unrealistic, given their extended structures. In reality, Gaia may detect multiple compact bright features in a single galaxy, and the centroiding accuracy for the galaxy will depend on the number and brightness of these features. Since the purpose here is to investigate whether these proper motion patterns are potentially detectable by Gaia, we adopted an optimistic approach pending a more detailed study of the centroiding accuracies for such objects. The simulation results must be interpreted keeping this assumption in mind. Clearly, increasing the redshift range will increase the number of objects with small angular diameter, but the amplitude of the proper motion patterns also decreases at higher redshift.

The cosmological proper motion was calculated according to the assumptions in Sect.~\ref{cos_prop_mot}. The proper motion components were then obtained by projecting this velocity on the tangent plane defined by the unit vector in the direction of the CMB antapex. In addition, we calculated random proper motion components of the galaxies by assuming a velocity dispersion of $v = 750$~km~s$^{-1}$ and again computed the proper motion components from the line-of-sight comoving distance.
These different effects were simulated together with the Galactocentric acceleration using Eq.~\eqref{Eq:sim1}, but with the modified values for the coefficients.

\subsection{Tools and other simulation parameters}
For our simulations we used AGISLab \citep{2012A&A...543A..15H}, which is a simulation tool used to help develop and test concepts and the corresponding algorithms for the astrometric global iterative solution (AGIS; \citealt{2012A&A...538A..78L}). AGISLab implements much of the functionality of the real Gaia data reduction software, but in addition is able to generate simulated observations using realistic estimates of the observation noise as a function of magnitude. Using this tool, sets of true and noisy astrometric parameters can be generated for a range of different types of sources distributed on the celestial sphere. In addition, attitude parameters for Gaia can be determined in the iterative least-squares solution (AGIS) using splines. For both the source and attitude parameters, different initial systematic and random noise values can be added and are then compared to the true values as an iterative solution proceeds until an acceptable level of convergence is achieved.  AGISLab generates the observations based on, for example, the satellite CCD geometry and its orbit. Additionally, the direction to a source is computed using the full Gaia relativity model \citep{2003AJ....125.1580K, 2004PhRvD..69l4001K}. A set of observation equations are used to construct the least-squares problem for the astrometric parameters using normal equations \citep{2012A&A...538A..78L}, which are then solved using a conjugate gradient algorithm described in \cite{2012A&A...538A..77B}. 
To the conventional features of AGISLab we added the option to include Markov chain photo-centre variability  as a perturbation to the transit times on the Gaia CCDs. Finally, AGISLab contains a number of utilities to generate statistics and graphical output.

In the simulations presented here, we used AGISLab without scaling down \citep[see][]{2012A&A...543A..15H} the mission and assumed an attitude-modelling knot interval of 30 seconds. For the starting noisy astrometric parameters we used 100~mas random errors and 10~mas systematic errors, while for the attitude parameters we assumed a nominal attitude error of 10~$\mu$as. For the accuracy of the reference frame we compared the final positions and proper motions with different realisations of their true values, assuming a random variation of 100~$\mu$as for VLBI\_QUASARS in position and 10~$\mu$as~yr$^{-1}$ in proper motion for all quasars. For the simulations we used an antapex direction in Galactic coordinates of ($l = 83.99^\circ$, $b = -48.26^\circ$). We assumed a Galactocentric acceleration of 4.3 $\mu$as yr$^{-1}$, CMB velocity of 369.0 km~s$^{-1}$, a Hubble constant of 67.3 km~s$^{-1}$~Mpc$^{-1}$, and a five-year mission.

\section{Determining the reference frame and proper motion patterns}\label{Sec:refframe} 
The relative measurement principle of Gaia results in astrometric parameters (positions and proper motions) that are determined with up to six degrees of freedom in the orientation $\vec{\epsilon} = (\epsilon_x,\ \epsilon_y,\ \epsilon_z)$ and the spin $\vec{\omega} = (\omega_x,\ \omega_y,\ \omega_z)$ of the Gaia reference frame relative to the ICRS at an adopted frame rotator epoch, $t_\text{fr}$, taken to be the same as the reference epoch of the astrometric parameters.  To express the final astrometric results in a celestial reference frame that closely matches the ICRS, the $\vec{\epsilon}$ and $\vec{\omega}$ parameters must be estimated from several sets of sources in a least-squares solution. The determined parameters can then be used to correct the reference frame to coincide with the ICRS. This frame rotation determination is made subsequent to AGIS using three different sets of sources:

\begin{itemize}
\item {\tiny VLBI\_QUASAR} \citep[called S$_{\rm P}$ in][]{2012A&A...538A..78L}~-- a subset consisting of the optical counterparts of a few thousand radio VLBI objects with known positions and proper motions in the ICRS independent of Gaia, which help to constrain $\vec{\epsilon}$. They are typically the optical counterparts of extragalactic objects with accurate positions 
from VLBI. These quasars can also be used to calculate the spin parameter ($\vec{\omega}$), but being very small in number, their contribution to the determination of spin is small.
\item {\tiny GAIA\_QUASAR} (S$_{\rm NR}$)~-- a larger subset consisting of hundreds of thousands of quasar-like objects ($\sim 10^{5}~$--$~10^{6}$) taken from ground based and photometric surveys that help to  constrain $\vec{\omega}$. These quasars do not have accurately known positions in ICRS and therefore cannot be used to calculate $\vec{\epsilon}$. They are assumed to define a kinematically non-rotating celestial frame.
\item {\tiny ICRS\_STAR} (S$_{\rm PM}$)~-- a subset of primary sources that have positions and/or proper motions that are accurately determined with respect to the ICRS independent of Gaia. This could include radio stars observed by radio VLBI interferometry, or stars whose absolute proper motions have been determined 
by some other means. 
\end{itemize}
For the present investigation we did not consider the subset {\tiny ICRS\_STAR} set, but added another subset of point-like sources at low redshift ($z < 0.03$): 
\begin{itemize} 
\item {\tiny GAIA\_GALAXY} -- a subset of point-like galactic nuclei sources with known redshift that may have measurable proper motion, albeit small. This subset can be used to probe redshift-dependent proper motion patters (see Sect.~\ref{cos_prop_mot}).
\end{itemize}
We determined the orientation and spin parameters together with the acceleration parameters, combined into a single parameter array 
$\vec{\theta}=[\epsilon_x~\epsilon_y~\epsilon_z~\omega_x~\omega_y~\omega_z ~\tilde{a}_x~\tilde{a}_y~\tilde{a}_z]'$  using a least-squares estimation of the positions and proper motions of subsets {\tiny GAIA\_QUASAR} and {\tiny VLBI\_QUASAR} in the two frames. For {\tiny GAIA\_QUASAR} the expression for the apparent proper motion in the Gaia reference frame is given by Eq.~(108) in \citet{2012A&A...538A..78L}, namely
\begin{align}
\mu_{\alpha*} & = \vec{q}'\vec{\omega} + \vec{p}'\vec{\tilde{a}} \\
\mu_{\delta}  & = -\vec{p}'\vec{\omega} + \vec{q}'\vec{\tilde{a,}} 
\end{align}
and the corresponding Eq.~(110) for {\tiny VLBI\_QUASAR} simplifies to
\begin{align}
\Delta\alpha* & = \vec{q}'\vec{\epsilon}\\
\Delta\delta  & = -\vec{p}'\vec{\epsilon} 
\end{align}
if the position differences are measured at the reference epoch $t_\text{ep}$. 
In these equations $\vec{p} = [-\sin\alpha,\ \cos\alpha,\ 0]'$ and $\vec{q} = [-\sin\delta\cos\alpha,\ -\sin\delta\sin\alpha,\ \cos\delta]'$. 

After the parameters $\vec{\epsilon}$, $\vec{\omega}$ and $\vec{\tilde{a}}$ were determined, the {\tiny GAIA\_GALAXY} subset of low-redshift galaxies can be used to estimate the redshift-dependent parameters\footnote{An alternative, elegant method for estimating these parameters could be to use vector spherical harmonics as outlined in \cite{2012A&A...547A..59M}. However, the Gaia data processing currently uses the above approach as baseline \citep{2012A&A...538A..78L}.} To determine the velocity of the solar system relative to the CMB, we must assume to know the Hubble constant and then solve in a  similar way for $\vec{\theta} = [v_x~v_y~v_z~\tilde{a}_x~\tilde{a}_y~\tilde{a}_z]'$, assuming $\vec{\epsilon}$, $\vec{\omega,}$ and $H_0$ are known. Conversely, if we assume that the velocity of the solar system relative to the CMB is known, then it is possible to determine $H_0$ by solving for $\vec{\theta}=[H_0~\tilde{a}_x~\tilde{a}_y~\tilde{a}_z]'$. We note that it is not necessary to solve for the acceleration term in each case, but the acceleration effect is also present when using low-redshift galaxies, and hence it is useful to include it as a consistency check and to obtain the correlation. For these secondary calculations\footnote{In principle it would be possible to directly combine the primary with a secondary calculation, but it practice the primary calculation is the baseline for the Gaia data processing, and it was decide to refrain from complicating this critical software.}, we cannot use the quasars because the effect rapidly decreases with increasing redshift and most quasars would contribute very little to the solution (see Sect.~\ref{Sec:sim:galaxies}).
\section{Results}\label{Sec:results}
To determine how well we recover the reference frame, we ran one hundred different realisations of the various simulation cases described below. The results of a comparison between the results and the true values gives the errors from which the mean and standard deviation values are found.  
\begin{table*}[htbp!]
\centering \caption[]{Summary of the different simulations. In all cases the nominal Gaia observation noise and a nominal attitude noise of 10 $\mu$as is used. Case A is an ideal reference simulation, while cases B and C introduce quasar variability with different timescales. 
In case D we use the IGQL data set to show the impact of a non-uniform sky distribution Fig.~\ref{Fig:skydistribution}.\label{Tab:summary}}
\begin{tabular}{ccccc}
\toprule\toprule
Data set & Case  &  Characteristic        & {\tiny VLBI\_} & {\tiny GAIA\_} \\
         &       &  time                  & {\tiny QUASAR} & {\tiny QUASAR} \\
         &       &  $\tau_{\rm cor}$ [yr] & \sv{}~[$\mu$as]& \sv{}~[$\mu$as]\\ \midrule[0.2pt]
Case A   & GUMS  &  {\rm --}              & 0              & 0              \\
Case B   & GUMS  &  2                     & 100            & 100            \\
Case C   & GUMS  &  10                    & 100            & 100            \\
Case D   & IGQL  &  {\rm --}              & 0              & 0              \\
\bottomrule
\end{tabular}
\end{table*}

Table \ref{Tab:summary} gives a summary of the different simulations. Case A is a reference data set where only Gaia nominal observation noise and the nominal attitude noise were added. Using the same noise assumptions, cases B and C then assessed the impact of adding quasar variability with different characteristic timescales. Finally, case D shows the results of using the ground-based quasar list IGQL (Sect.~\ref{Sec:sim}) under nominal conditions similar to case A to determine the impact of using a non-uniform sky distribution. 

In each of these cases, a first simulation was made using quasars to accurately determine the reference frame parameters, $\vec{\epsilon}$ and $\vec{\omega}$, and the acceleration of the solar system, $\vec{\tilde{a}}$. In two independent secondary simulations we
then used the values of $\vec{\epsilon}$ and $\vec{\omega}$ found using the quasars to determine either the instantaneous velocity of the solar system relative to the CMB, or the Hubble constant using low-redshift galaxies as described in Sect.~\ref{Sec:refframe}. As these two parameters are degenerate, we cannot solve for both at the same time and hence we had two secondary simulations. For the secondary galaxy simulations no variability was assumed, and we also calculated the acceleration of the solar system, $\vec{\tilde{a}}$, which is less accurate because we now used fewer sources. The assumption of using fewer sources may be pessimistic given that the number of objects detected by Gaia is much higher, but a detailed study of the centroiding accuracy for galaxies would be needed for a more realistic simulation of this effect.

\begin{table*}[h]
        \footnotesize
        \centering
        \caption[]{Results of the simulations. For each case in Table~\ref{Tab:summary}, we give the mean value (Mean) and standard deviation (Std) of 100 experiments, together with the average formal standard error ($\langle\sigma\rangle$) of the parameter. $x$, $y$, $z$ are the Galactic components of the vectors. \label{Tab:acceleration}}
        \begin{tabular}{ccccccccccc}
                \toprule\toprule
                & \multicolumn{1}{c}{$$}
                & \multicolumn{3}{c}{Orientation ($\vec{\epsilon}$)}
                & \multicolumn{3}{c}{Spin ($\vec{\omega}$)}
                & \multicolumn{3}{c}{Acceleration ($\vec{\tilde{a}}$)} \\
                & \multicolumn{1}{c}{$$}
                & \multicolumn{3}{c}{[$\mu$as]}
                & \multicolumn{3}{c}{[$\mu$as~yr$^{-1}$]}
                & \multicolumn{3}{c}{[$\mu$as~yr$^{-1}$]} \\
                \cmidrule[0.2pt](lr){3-5}       
                \cmidrule[0.2pt](lr){6-8} 
                \cmidrule[0.2pt](lr){9-11}
                & $ $ & $x$ & $y$ & $z$ & $x$ & $y$ & $z$ & $x$ & $y$ & $z$ \\ 
                \cmidrule[0.2pt](lr){3-5}       
                \cmidrule[0.2pt](lr){6-8} 
                \cmidrule[0.2pt](lr){9-11}
                \multicolumn{1}{c}{\multirow{1}{*}{Case A}} &
                \multicolumn{1}{c}                              {Mean}                         &       -0.249  &       -0.234  &       -0.126  &       0.001   &       -0.004  &       -0.003  &       4.304   &       0.003   &       -0.026 \\
                \multicolumn{1}{c}{}        &   {Std}                   &       3.282   &       3.425   &       3.069   &       0.125   &       0.158   &       0.166   &       0.149   &       0.144   &       0.157 \\
                \multicolumn{1}{c}{}&{$\langle\sigma\rangle$}   &       2.211   &       2.254   &       2.445   &       0.144   &       0.147   &       0.159   &       0.145   &       0.146   &       0.159 \\ 
                \\
                \cmidrule[0.2pt](lr){3-5}       
                \cmidrule[0.2pt](lr){6-8} 
                \cmidrule[0.2pt](lr){9-11}
                \multicolumn{1}{l}{\multirow{1}{*}{Case B}} &
                \multicolumn{1}{c}                              {Mean}                         &       0.169   &       0.182   &       0.428   &       -0.003  &       0.001   &       0.007   &       4.308   &       0.002   &       -0.007 \\
                \multicolumn{1}{c}{}        &   {Std}                   &       3.909   &       3.695   &       4.329   &       0.145   &       0.182   &       0.167   &       0.162   &       0.152   &       0.175 \\

                \multicolumn{1}{c}{}&{$\langle\sigma\rangle$}   &       2.239   &       2.284   &       2.473   &       0.154   &       0.156   &       0.169   &       0.154   &       0.155   &       0.169 \\ 
                \\
                \cmidrule[0.2pt](lr){3-5}       
                \cmidrule[0.2pt](lr){6-8} 
                \cmidrule[0.2pt](lr){9-11}
                \multicolumn{1}{c}{\multirow{1}{*}{Case C}} &
                \multicolumn{1}{c}                              {Mean}                         &       0.216   &       -0.072  &       -0.148  &       -0.014  &       0.013   &       0.016   &       4.307   &       0.009   &       -0.001 \\
                \multicolumn{1}{c}{}        &   {Std}                   &       4.605   &       4.281   &       4.469   &       0.136   &       0.178   &       0.171   &       0.154   &       0.146   &       0.165 \\
                \multicolumn{1}{c}{}&{$\langle\sigma\rangle$}   &       2.260   &       2.304         &       2.494   &       0.148   &       0.150   &       0.163   &       0.148   &       0.150   &       0.163 \\ 
                \\
                \cmidrule[0.2pt](lr){3-5}       
                \cmidrule[0.2pt](lr){6-8} 
                \cmidrule[0.2pt](lr){9-11}
                \multicolumn{1}{c}{\multirow{1}{*}{Case D}} &
                \multicolumn{1}{c}                              {Mean}                         &       3.818   &       0.633   &       1.427   &       -1.083  &       -4.473  &       1.184   &       5.578   &       1.657   &       -1.654 \\
                \multicolumn{1}{c}{}        &   {Std}                   &       3.217   &       3.779   &       4.559   &       0.151   &       0.165   &       0.185   &       0.175   &       0.151   &       0.203 \\
                \multicolumn{1}{c}{}&{$\langle\sigma\rangle$}   &       2.157   &       2.136   &       2.838   &       0.171   &       0.167   &       0.211   &       0.172   &       0.167   &       0.211 \\ 
                \\
                \bottomrule
        \end{tabular}
\end{table*}
Table \ref{Tab:acceleration} shows the results of the reference frame determination based on quasars for cases A to D. For the reference case A the mean value for Galactocentric acceleration, which is determined simultaneously with the reference frame, is $\left|\vec{\tilde{a}}\right|=4.304~\mu$as~yr$^{-1}$ with a standard deviation and mean formal error of $\sim$$0.26~\mu$as~yr$^{-1}$ compared to the simulated input value of 4.3~$\mu$as~yr$^{-1}$. This shows that under optimal conditions, the Galactocentric acceleration can be determined to a few percent. The formal errors ($\langle\sigma\rangle$) typically underestimate the standard deviation of the actual errors (Std) because of the source variability and the attitude noise, which are not included in the formal errors. The mean values are consistent with the standard deviations except in case D discussed below. The general trend is as expected when comparing cases A to D, as each case adds more complexity and the standard deviations and the mean formal errors all increase slightly. 

The addition of photo-centre variability in cases B and C with characteristic timescales of 2 and 10 years, respectively, does not significantly degrade the solution for Galactocentric acceleration. The mean value for Galactocentric acceleration is very similar at $\left|\vec{\tilde{a}}\right|=4.308~\mu$as~yr$^{-1}$, but both the standard deviation and the mean formal errors of these runs only increase very modestly and roughly agree with, but are slightly better than, the predictions of \cite{2014ApJ...789..166P}. These differences can be partially explained by the use of different numbers of quasars in the two estimates -- 170\,000 in theirs and 500\,000 in ours. 

\begin{figure*}[htbp!]
\includegraphics[width = \columnwidth]{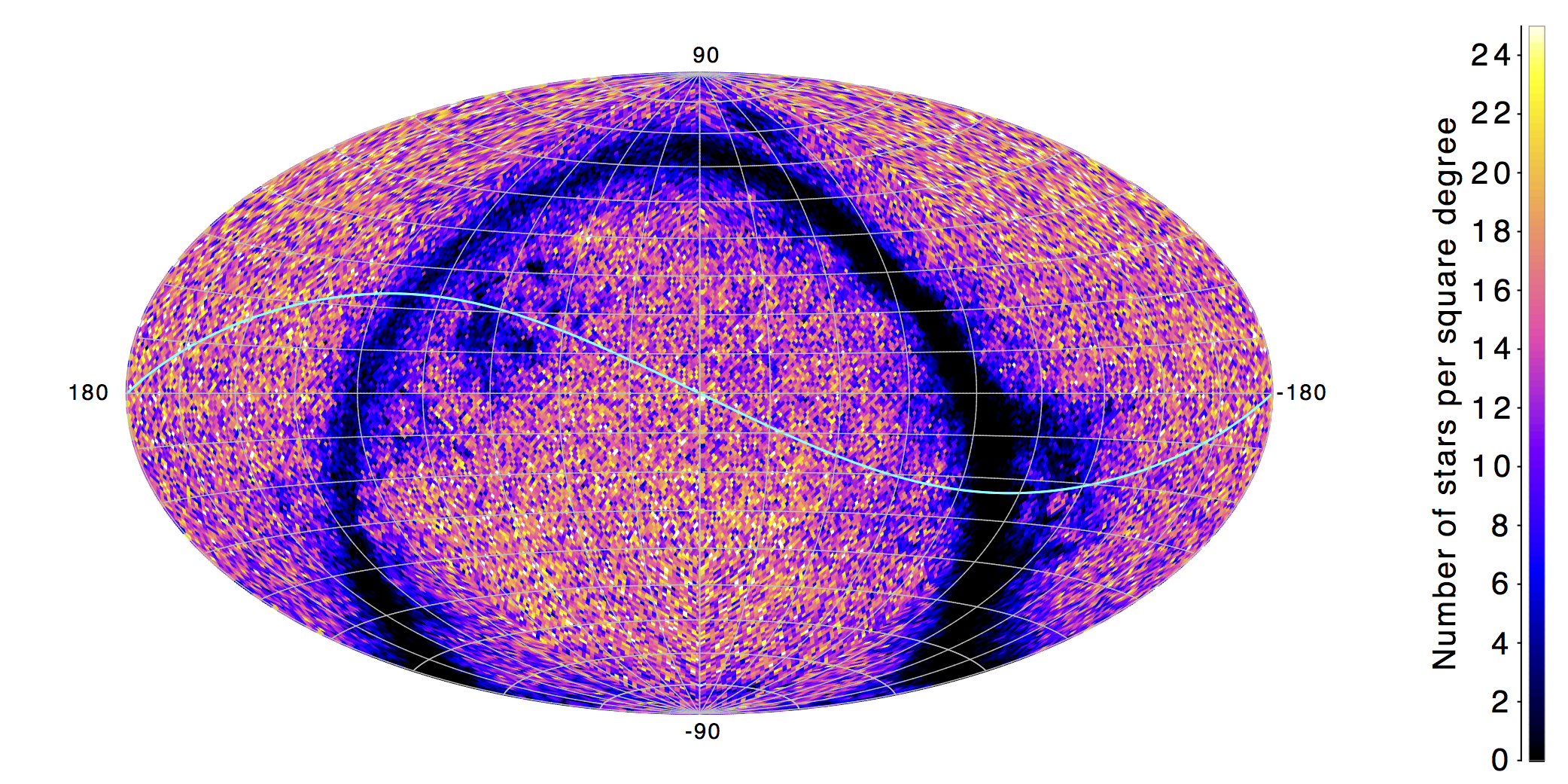}
\includegraphics[width = \columnwidth]{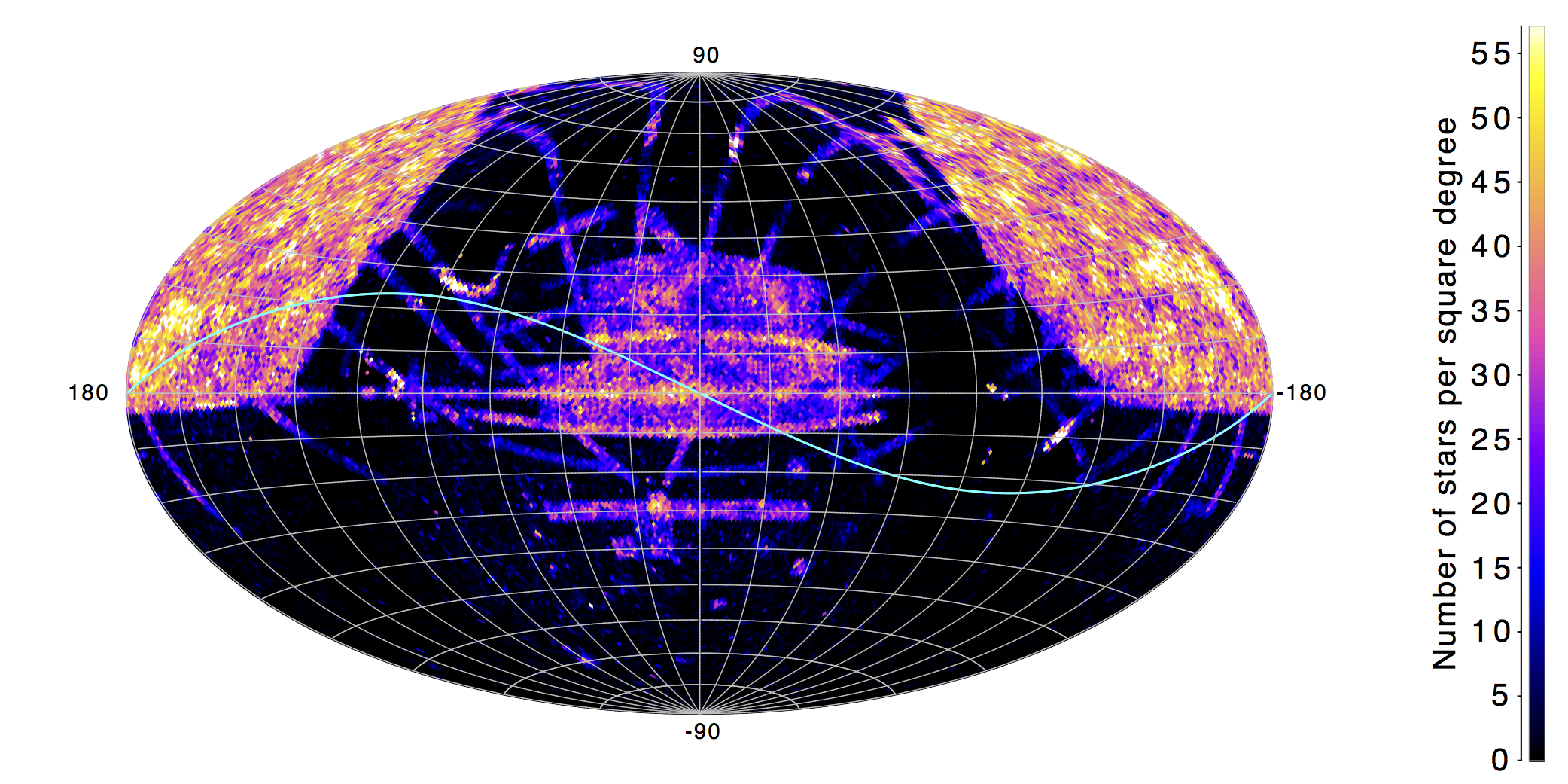}
\caption{All-sky maps showing the median number of sources per pixel in an equatorial Hammer-Aitoff projection. {\em Left:} The quasar distribution on the sky from the GUMS catalogue. {\em Right:} The quasar distribution on the sky from the IGQL catalogue. The cyan line denotes the ecliptic.\label{Fig:skydistribution}}
\end{figure*}

In addition to using the GUMS quasar data, which are well distributed on the sky apart from the Galactic plane, we also used the IGQL dataset in case D. The results for acceleration are poorer (for the IGQL we derive a value of $\left|\vec{\tilde{a}}\right| = 6.049~\mu$as~yr$^{-1}$ with significant components directed away from the Galactic centre), but the standard deviations and formal errors do not change considerably from Case A. The reason for different mean value is the non-uniform sky distribution of the data set, particularly in the southern hemisphere, where the data set is rather sparse (see Fig.~\ref{Fig:skydistribution}) compared to the GUMS catalogue. This results in the spin parameter ($\vec{\omega}$) being poorly determined for the reference frame, and this consequently  affects the determination of the acceleration vector. By comparing the resulting correlation matrices for GUMS and IGQL, we show  in Table \ref{Tab:correlation}, highlighted in red, that a weak correlation exists between the components of $\vec{\omega}$ and $\vec{\tilde{a}}$ in the GUMS catalogue, but this is much stronger in the IGQL data set. The IGQL will be used in the early Gaia processing and care should be taken when interpreting the reference frame and Galactocentric acceleration results. However, as Gaia detects more quasars itself, the sparse distribution found in the IGQL catalogue will be filled in and the results presented here based on the GUMS catalogue should become more representative. We note that many of the elements of the correlation matrix are zero, which results purely from our choice of reference frame epoch and the source reference epochs, which are all the same. Choosing different reference times would result in non-zero entries in the correlation matrix, but for comparison purposes, the current choice is more useful.  

\begin{table*}[htbp!]
        \footnotesize
        \centering
        \caption[]{Symmetric correlation matrix calculated from the frame rotation ($\vec{\epsilon}$, $\vec{\omega}$) and acceleration ($\vec{\tilde{a}}$) parameters for a GUMS dataset (upper right triangle) to be compared with an IGQL dataset (lower left triangle). The main differences in the IGQL results are highlighted in red. There is a significantly stronger correlation between $\vec{\omega}$ and $\vec{\tilde{a,}}$ which leads to poorer results. \label{Tab:correlation}}
        \begin{tabular}{cccccccccc}
                \toprule
                & $\epsilon_x$ & $\epsilon_y$ & $\epsilon_z$ & $\omega_x$ & $\omega_y$ & $\omega_z$ & $\tilde{a}_x$ & $\tilde{a}_y$ & $\tilde{a}_z$ \\
                \toprule
                $\epsilon_x$ &{\bf 1.0}&  0.001  & -0.025  &  0.000                     &  0.000            &  0.000            &  0.000            &  0.000            &  0.000 \\
                $\epsilon_y$ &  0.038  &{\bf 1.0}&  0.017  &  0.000                     &  0.000            &  0.000            &  0.000            &  0.000            &  0.000 \\
                $\epsilon_z$ & -0.037  & -0.045  &{\bf 1.0}&  0.000                     &  0.000            &  0.000            &  0.000            &  0.000            &  0.000 \\
                \cmidrule[0.2pt](lr){2-10}
                $\omega_x$   &  0.000  &  0.000  &  0.000  &{\bf 1.0}          &  0.003            & -0.021            &  0.000            &  0.010            &  0.010 \\
                $\omega_y$   &  0.000  &  0.000  &  0.000  &{\color{red} 0.106}&{\bf 1.0}          &  0.011            & -0.010            &  0.000            & -0.048 \\
                $\omega_z$   &  0.000  &  0.000  &  0.000  &{\color{red} 0.071}&{\color{red}-0.182}&{\bf 1.0}          & -0.010            &  0.049            &  0.000 \\
                \cmidrule[0.2pt](lr){2-10}
                $\tilde{a}_x$        &  0.000  &  0.000  &  0.000  &  0.002            &{\color{red}-0.444}&{\color{red} 0.328}&{\bf 1.0}          &  0.011            & -0.016 \\
                $\tilde{a}_y$        &  0.000  &  0.000  &  0.000  &{\color{red} 0.444}& -0.004            &{\color{red} 0.264}&{\color{red} 0.118}&{\bf 1.0}          &  0.001 \\
                $\tilde{a}_z$        &  0.000  &  0.000  &  0.000  &{\color{red}-0.330}&{\color{red}-0.262}&  0.002            &{\color{red} 0.078}&{\color{red}-0.189}&{\bf 1.0} \\
                \bottomrule
        \end{tabular}
\end{table*}

\begin{table*}[h]
        \footnotesize
        \centering
        \caption[]{Results from simulations using the GUMS galaxy catalogue for all cases. See text for details. The two estimates of the acceleration
based on galaxies gave very similar results (to within a few percent), therefore we only list those from the first case.
                \label{Tab:velocityHubble}}
        \begin{tabular}{cccccccccccc}
                \toprule\toprule
                & \multicolumn{1}{c}{$ $}
                & \multicolumn{3}{c}{Acceleration ($\vec{\tilde{a}}$)} 
                & \multicolumn{4}{c}{Velocity ($\vec{v}$)}
                & \multicolumn{2}{c}{Direction}
                & \multicolumn{1}{c}{Hubble constant} \\
                & \multicolumn{1}{c}{$ $}
                & \multicolumn{3}{c}{[$\mu$as~yr$^{-1}$]} 
                & \multicolumn{4}{c}{[km~s$^{-1}$]}
                & \multicolumn{2}{c}{[$^\circ$]}
                & \multicolumn{1}{c}{[km~s$^{-1}$~Mpc$^{-1}$]} \\
                \cmidrule[0.2pt](lr){3-5}       
                \cmidrule[0.2pt](lr){6-9}       
                \cmidrule[0.2pt](lr){10-11} 
                \cmidrule[0.2pt](lr){12-12}
                & $ $ & x & y & z & x & y & z &$\left|\vec{v}\right|$& l & b & $H_0$\\
                \cmidrule[0.2pt](lr){3-5}       
                \cmidrule[0.2pt](lr){6-9}       
                \cmidrule[0.2pt](lr){10-11} 
                \cmidrule[0.2pt](lr){12-12}
                \multicolumn{1}{c}{\multirow{1}{*}{Case A}}   &
                \multicolumn{1}{c}                        {Mean}                  &   4.342      &       0.029   &       -0.005  &   20.053      &       234.445 & -273.981        &   361.154     &   85.111      &       -49.343 &   65.748      \\
                \multicolumn{1}{c}{}            & {Std}                   &   0.241      &       0.195   &       0.234   &   65.904      &       55.394  &       64.385  &   107.505    &   $$          &       $$              &   11.722      \\
                \multicolumn{1}{c}{}& {$\langle\sigma\rangle$}&   0.270  &       0.269   &       0.270   &   98.403      &       97.522  &       96.899  &   169.066    &   $$          &   $$          &   18.034      \\
                \\
                \cmidrule[0.2pt](lr){3-5}       
                \cmidrule[0.2pt](lr){6-9}       
                \cmidrule[0.2pt](lr){10-11} 
                \cmidrule[0.2pt](lr){12-12}
                \multicolumn{1}{c}{\multirow{1}{*}{Case B}}   &
                \multicolumn{1}{c}                        {Mean}                  &   4.300      &       0.016   &  -0.017       &   27.487      &       238.212 & -270.408        &   361.415&   83.417   &       -48.434 &   65.895      \\
                \multicolumn{1}{c}{}            & {Std}                   &   0.214      &       0.217   &       0.211   &   53.767      &       56.126  &       61.535  &    99.134&   $$              &   $$          &   10.991      \\
                \multicolumn{1}{c}{}& {$\langle\sigma\rangle$}&   0.271  &       0.270   &       0.270   &   98.665      &       97.796  &       97.166  &   169.529&   $$              &   $$          &   18.153      \\
                \\
                \cmidrule[0.2pt](lr){3-5}       
                \cmidrule[0.2pt](lr){6-9}       
                \cmidrule[0.2pt](lr){10-11} 
                \cmidrule[0.2pt](lr){12-12}
                \multicolumn{1}{c}{\multirow{1}{*}{Case C}}   &
                \multicolumn{1}{c}                        {Mean}                  &   4.299      &       0.016   &       -0.018  &   27.524      &       238.205 & -270.348        &   361.369&   83.409   &       -48.428 &   65.887      \\
                \multicolumn{1}{c}{}            & {Std}                   &   0.214      &       0.217   &       0.211   &   53.758      &       56.108  &       61.481  &    99.086&   $$              &   $$          &   10.984      \\
                \multicolumn{1}{c}{}& {$\langle\sigma\rangle$}&   0.271  &       0.270   &       0.270   &   98.665      &       97.796  &       97.166  &   169.529&   $$              &       $$              &   18.153      \\
                \\
                \cmidrule[0.2pt](lr){3-5}       
                \cmidrule[0.2pt](lr){6-9}       
                \cmidrule[0.2pt](lr){10-11} 
                \cmidrule[0.2pt](lr){12-12}
                \multicolumn{1}{c}{\multirow{1}{*}{Case D}}   &
                \multicolumn{1}{c}                        {Mean}                  &   4.328      &       0.045   &       0.194   &   17.152      &       238.202 & -299.719        &   383.231     &   85.881      &       -51.452 &   69.711      \\
                \multicolumn{1}{c}{}            & {Std}                   &   0.219      &       0.221   &       0.208   &   60.978      &       55.090  &       57.167  &   100.106    &   $$          &   $$          &   11.190      \\
                \multicolumn{1}{c}{}& {$\langle\sigma\rangle$}&   0.275 &       0.274   &       0.275   &  100.173      &       99.303  &       98.691  &   172.150    &   $$          &   $$          &   18.053      \\
                \\
                \bottomrule
        \end{tabular}
\end{table*}
In Table \ref{Tab:velocityHubble} the results from simulations using low-redshift galaxies from the GUMS catalogue are presented. We first determine the individual reference frame parameters $\vec{\epsilon}$ and $\vec{\omega}$, using the quasar simulations tabulated in Table \ref{Tab:acceleration}, which are then fixed input for theses secondary simulations. Two independent secondary simulations were made. First, we determined the instantaneous velocity of the solar system  with respect to the CMB together with the acceleration. Second, the Hubble constant was solved together with the acceleration. It is not necessary to make these two estimates of the acceleration based on low-redshift galaxies, but doing so allows the correlation with the parameters of interest to be estimated, and it provides a consistency check with the value determined from the quasar data.

If we compare the results for the acceleration determined from the low-redshift galaxies with that determined from the quasars in Table \ref{Tab:acceleration}, we see that the actual values do not change significantly. However, the standard deviation and mean formal errors have increased by factors of $\sim$1.5 and $\sim$1.8, respectively, for Case A, which is slightly lower than the expected value of $\sqrt{5}$ due to the difference in the number of objects used. This discrepancy is probably a result
of the lack of variability in the galaxy observations and the somewhat higher brightness of the galaxies compared with the quasars (the number of galaxies peaks at $\sim$17 G magnitude -- see Fig.~\ref{Fig:weightGalaxies} bottom left). Little information on the centroiding accuracy of point-like galaxies (or components thereof) is currently available, and our values here certainly underestimate the real errors. The differences of the acceleration determined from the two secondary simulations are well within the expected variance, and thus we have only reported those from the velocity case.
In addition, for case D, the acceleration results are much better. This is because case D in Table \ref{Tab:acceleration} used the IGQL quasar dataset, with a non-uniform sky distribution, to estimate the values of $\vec{\epsilon}$, $\vec{\omega,}$ and $\vec{\tilde{a,}}$ while in Table \ref{Tab:velocityHubble} we use the uniformally distributed galaxies to estimate the acceleration, which now gives better results. Clearly, the sky distribution is important.

The results for the instantaneous velocity of the solar system with respect to the CMB are also shown in Table~\ref{Tab:velocityHubble}. The recovered components of the velocity, magnitude, and direction are given. In the simulations we used an apex direction of ($l = 263.99^\circ$, $b = 48.26^\circ$) and a magnitude of $\left|\vec{v}\right| = 369~{\rm km~s}^{-1}$. The recovered mean values and direction agree well (within five percent) in all simulations. The magnitude of the standard deviation is just over 100~km~s$^{-1}$, while the magnitude of the mean formal error is of the order of 170~km~s$^{-1}$. These errors are relatively large, but we recall that we used only 100\,000 objects in the redshift range $z < $ 0.03, while the largest contribution to this effect is dominated by a small number of objects in the range $z <$ 0.01 (see Fig.~\ref{Fig:weightGalaxies} right).
The results given in Table \ref{Tab:velocityHubble} for case D are slightly poorer because the fixed reference frame parameters $\vec{\epsilon}$ and $\vec{\omega}$ were less accurately determined using the IGQL quasars. They therefore also have a weak effect
on determining the proper motion patterns in these secondary simulations. 

For the Hubble constant the recovered mean values also agree
well (to within four percent) in all simulations. Again, the errors are relatively large, the standard deviation is  11-12~km~s$^{-1}$~Mpc$^{-1}$, while the mean formal error is of the order of 18~km~s$^{-1}$~Mpc$^{-1}$. Clearly, detecting more sources in the low-redshift range $z <$ 0.01 would greatly improve these results, but it remains unclear for now how many more objects could be used for this determination and how well these extended objects can be centroided. We here used a very optimistic assumption concerning the centroiding accuracy, which means that measuring these proper motion patterns from galaxies will remain a challenging task.

\section{Conclusions}\label{Sec:conclusions}

We have assessed how well Gaia can improve the optical realisation of ICRS using sets of quasars to define a kinematically stable non-rotating reference frame with the barycentre of the solar system as its origin. Photo-centric variability of the quasars was simulated using a Markov chain model at the level of 100~$\mu$as and does not appear to significantly perturb the results for either the reference frame determination or the proper motion pattern that is due to the acceleration of the solar system. 

The initial quasar list (IGQL) does not have an ideal sky distribution to determine the reference frame or proper motion patterns, and care should be taken in early Gaia data processing when interpreting theses results. However, as Gaia detects more quasars itself, the sky distribution will improve and eventually become more uniform and representative of the GUMS catalogue used in these simulations. It should be noted that other surveys exist, such as the WISE project \citep[e.g.,][]{2015ApJS..221...12S}, which could improve the situation in the early Gaia solutions.

We reviewed the various proper motion effects that could affect the Gaia data processing, and we tried to quantify their respective magnitudes. Most of the effects are either not detectable by astrometry or do not seem to be strong enough to be reliably detected by Gaia. However, a cosmological proper motion pattern that is due to the instantaneous velocity of the solar system with respect to the CMB might be just about measurable. In contrast to the acceleration of the solar system, this cosmological proper motion pattern is redshift dependent and only significant at low redshift. Thus it cannot be determined from the more distant quasars, and instead low-redshift point-like galaxies must be used. This source type has not yet been considered in the baseline processing for the Gaia mission.

From this proper motion pattern our velocity relative to the CMB can be determined provided we assume that the Hubble constant is known. Alternatively, if we assume our velocity is know from other missions, such as Planck \citep{2014A&A...571A..27P}, we can estimate the Hubble constant from this pattern of proper motions. We find that both measurements might be just within the reach of Gaia, provided that a suitable set of low-redshift point-like objects can be identified and used in the processing. The onboard detection and data processing (centroiding) of such low-redshift point-like objects will ultimately limit whether this effect can be measured in practice. 

\begin{acknowledgements}
The authors thank the Swedish National Space Board (SNSB) for financial support, without which this project would not have been possible. We also thank the referee, V.~Makarov, for useful comments and suggestions on the manuscript.
\end{acknowledgements}

\appendix

\section{Statistical proper motion of quasars that are due to the random motion of lensing objects.} \label{Sec:appB}

\begin{figure}[h]
        \vspace{0.0cm} \centering
        \includegraphics[width = 2.5in]{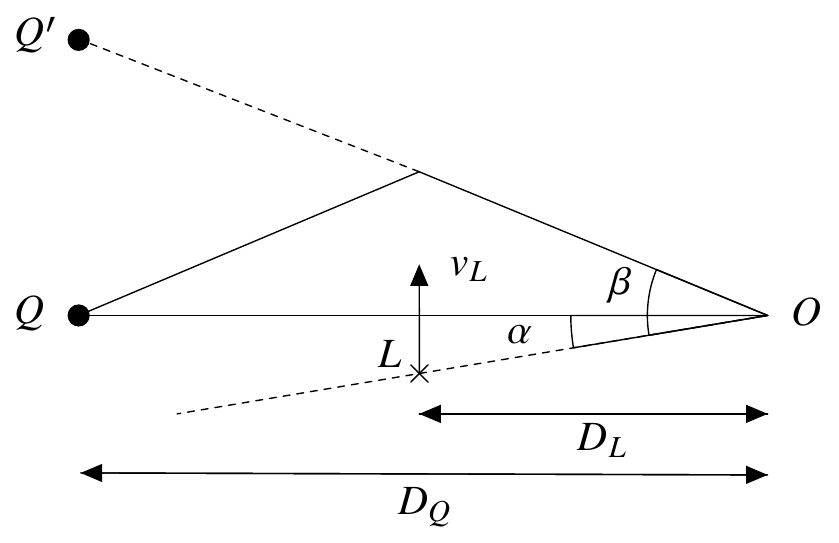}
        \caption{Lensing parameters.\label{Fig:lens}}
\end{figure}

Consider a point source $Q$ and a lensing object $L$ at a distance $D_Q$ and $D_L$  from an observer $O$ (see Fig. \ref{Fig:lens}). Let $\angle{QOL} = \alpha$ and $\angle{Q'OL} = \beta$, where $Q'$ is the shifted image of source $Q$ that is due to the presence of the lens. Then the lens equation is given by \citep{2008cosm.book.....W}
\begin{equation}\label{a1}
\beta(\beta - \alpha) = \theta^2_E = \frac{4GM_L}{c^2}\left(\frac{1}{D_L} - \frac{1}{D_Q} \right)\,,
\end{equation}
where $\theta_E$ is the Einstein radius. We assume that the point source is a quasar so that we can assume it to be at rest. Then the proper motion is $\mu_Q = 0$. If the lens has a low velocity $v_L$ along the direction perpendicular to the line of sight, then the proper motion of lens is given by
\begin{equation*}
\mu_L = \frac{v_L}{D_L} = -\frac{d\alpha}{dt}
\end{equation*}
The negative sign is chosen so that the lens motion and the image shift $Q'$ are in the same direction. If the lens is moving, then the image $Q'$ will also be moving, resulting in an apparent proper motion 
\begin{equation*}
\mu_{Q'} = \frac{d(\beta - \alpha)}{dt} = \frac{d\beta}{dt} - \frac{d\alpha}{dt} = \frac{d\beta}{dt} + \mu_L
\end{equation*}
From Eq.~\eqref{a1}
\begin{equation*}
\beta^2 - \beta\alpha - \theta^2_E = 0\, .
\end{equation*}
The roots of this equation are given by
\begin{equation*}
\beta = \frac{\alpha}{2} \pm \sqrt{\frac{\alpha^2}{4} + \theta^2_E}
\end{equation*}
and differentiating with respect to time gives
\begin{equation*}
\frac{d\beta}{dt} = \frac{d\alpha}{dt}\left[\frac{1}{2} \pm \frac{1}{2}\left(1 + \frac{4\theta^2_E}{\alpha^2} \right)^{-1/2} \right]\, .
\end{equation*}
Assuming that $\alpha \gg \theta_E$, and taking the positive sign for an image in one direction only, we obtain
\begin{equation}\label{a2}
\mu_{Q'} \approx \mu_L\left(\frac{\theta_E}{\alpha} \right)^2\, .
\end{equation}
To calculate the combined statistical effects of many lenses at different distances, we first consider a shell at a certain mean distance from us. All the lenses in this shell are assumed to have the same mass and therefore have the same Einstein radius. Let $N$ be the projected density of the lenses in this shell. The mean-squares from Eq.~\eqref{a2} is then
\begin{equation*}
\left\langle\mu_Q'^2 \right\rangle = \left\langle\mu_L^2 \left(\frac{\theta_E}{\alpha} \right)^4 \right\rangle = \left\langle\mu_L^2 \right\rangle \left\langle \left(\frac{\theta_E}{\alpha} \right)^4 \right\rangle \, ,
\end{equation*}
where the second equality follows from the statistical independence between $\mu_L$ and $\alpha$. With source $Q$ at the centre, we draw a circular ring of radius $\alpha$ and thickness $d\alpha$. Then the ring will contain $2\pi\alpha N \,d\alpha$ lenses and the expectation value of $\alpha^{-4}$ is given by
\begin{align*}
E\left(\alpha^{-4}\right) = \int\limits_{\theta_E}^\infty{2\pi\alpha}N\frac{1}{\alpha^4}~d\alpha = \pi\theta_E^{-2}N \, .
\end{align*}
Therefore,
\begin{equation}\label{a3}
\left\langle\mu_Q'^2 \right\rangle  = \left\langle\mu_L^2 \right\rangle\tau \, ,
\end{equation}
where
\begin{align*}
\tau \equiv \pi\theta_E^2N
\end{align*}
is the probability that $Q$ will be inside the Einstein ring of some lens in the shell, also known as the photometric optical depth of the shell. Summing the contributions from all the shells, we see that Eq.~\eqref{a3} is valid also for the total effect, if $\tau$ is the total optical depth. We finally arrive at
\begin{equation}\label{a4}
\left(\mu_{Q'} \right)_{\text{RMS}} = \left(\mu_{L} \right)_{\text{RMS}}\times \sqrt{\tau}
.\end{equation}
 
\bibliographystyle{aa} 
\bibliography{agis}

\begin{thebibliography}{65}
\expandafter\ifx\csname natexlab\endcsname\relax\def\natexlab#1{#1}\fi

\bibitem[{{Andrei} {et~al.}(2012){Andrei}, {Anton}, {Barache}, {Bouquillon},
  {Bourda}, {Le Campion}, {Charlot}, {Lambert}, {Pereira Osorio}, {Souchay},
  {Taris}, {Assafin}, {Camargo}, {da Silva Neto}, \& {Vieira
  Martins}}]{2012sf2a.conf...61A}
{Andrei}, A.~H., {Anton}, S., {Barache}, C., {et~al.} 2012, in SF2A-2012:
  Proceedings of the Annual meeting of the French Society of Astronomy and
  Astrophysics, ed. S.~{Boissier}, P.~{de Laverny}, N.~{Nardetto}, R.~{Samadi},
  D.~{Valls-Gabaud}, \& H.~{Wozniak}, 61--66

\bibitem[{{Andrei} {et~al.}(2009){Andrei}, {Souchay}, {Zacharias}, {Smart},
  {Vieira Martins}, {da Silva Neto}, {Camargo}, {Assafin}, {Barache},
  {Bouquillon}, {Penna}, \& {Taris}}]{2009A&A...505..385A}
{Andrei}, A.~H., {Souchay}, J., {Zacharias}, N., {et~al.} 2009, \aap, 505, 385

\bibitem[{{Bailer-Jones} {et~al.}(2008){Bailer-Jones}, {Smith}, {Tiede},
  {Sordo}, \& {Vallenari}}]{2008MNRAS.391.1838B}
{Bailer-Jones}, C.~A.~L., {Smith}, K.~W., {Tiede}, C., {Sordo}, R., \&
  {Vallenari}, A. 2008, \mnras, 391, 1838

\bibitem[{{Bastian}(1995)}]{1995ESASP.379...99B}
{Bastian}, U. 1995, in ESA Special Publication, Vol. 379, Future Possibilities
  for bstrometry in Space, ed. M.~A.~C. {Perryman} \& F.~{van Leeuwen}, 99

\bibitem[{{Belokurov} \& {Evans}(2002)}]{2002MNRAS.331..649B}
{Belokurov}, V.~A. \& {Evans}, N.~W. 2002, \mnras, 331, 649

\bibitem[{{Boehringer} {et~al.}(1996){Boehringer}, {Neumann}, {Schindler}, \&
  {Kraan-Korteweg}}]{1996ApJ...467..168B}
{Boehringer}, H., {Neumann}, D.~M., {Schindler}, S., \& {Kraan-Korteweg}, R.~C.
  1996, \apj, 467, 168

\bibitem[{{Bombrun} {et~al.}(2012){Bombrun}, {Lindegren}, {Hobbs}, {Holl},
  {Lammers}, \& {Bastian}}]{2012A&A...538A..77B}
{Bombrun}, A., {Lindegren}, L., {Hobbs}, D., {et~al.} 2012, \aap, 538, A77

\bibitem[{{Book} \& {Flanagan}(2011)}]{2011PhRvD..83b4024B}
{Book}, L.~G. \& {Flanagan}, {\'E}.~{\'E}. 2011, \prd, 83, 024024

\bibitem[{Chatfield(2013)}]{chatfield2013analysis}
Chatfield, C. 2013, The Analysis of Time Series: An Introduction, Sixth
  Edition, Chapman \& Hall/CRC Texts in Statistical Science (Taylor \& Francis)

\bibitem[{{Doob}(1942)}]{1942AnMat..43..351D}
{Doob}, J.~L. 1942, Annals of Mathematics, 43, 351

\bibitem[{{ESA}(1997)}]{1997ESASP1200.....E}
{ESA}, ed. 1997, ESA Special Publication, Vol. 1200, {The HIPPARCOS and TYCHO
  catalogues. Astrometric and photometric star catalogues derived from the ESA
  HIPPARCOS Space Astrometry Mission}

\bibitem[{{Feng} \& {Bailer-Jones}(2013)}]{2013ApJ...768..152F}
{Feng}, F. \& {Bailer-Jones}, C.~A.~L. 2013, \apj, 768, 152

\bibitem[{{Fey} {et~al.}(2015){Fey}, {Gordon}, {Jacobs}, {Ma}, {Gaume},
  {Arias}, {Bianco}, {Boboltz}, {B{\"o}ckmann}, {Bolotin}, {Charlot},
  {Collioud}, {Engelhardt}, {Gipson}, {Gontier}, {Heinkelmann}, {Kurdubov},
  {Lambert}, {Lytvyn}, {MacMillan}, {Malkin}, {Nothnagel}, {Ojha},
  {Skurikhina}, {Sokolova}, {Souchay}, {Sovers}, {Tesmer}, {Titov}, {Wang}, \&
  {Zharov}}]{2015AJ....150...58F}
{Fey}, A.~L., {Gordon}, D., {Jacobs}, C.~S., {et~al.} 2015, \aj, 150, 58

\bibitem[{{Fouqu{\'e}} {et~al.}(2001){Fouqu{\'e}}, {Solanes}, {Sanchis}, \&
  {Balkowski}}]{2001A&A...375..770F}
{Fouqu{\'e}}, P., {Solanes}, J.~M., {Sanchis}, T., \& {Balkowski}, C. 2001,
  \aap, 375, 770

\bibitem[{{Gavazzi} {et~al.}(2009){Gavazzi}, {Adami}, {Durret}, {Cuillandre},
  {Ilbert}, {Mazure}, {Pell{\'o}}, \& {Ulmer}}]{2009A&A...498L..33G}
{Gavazzi}, R., {Adami}, C., {Durret}, F., {et~al.} 2009, \aap, 498, L33

\bibitem[{{Ghez} {et~al.}(2008){Ghez}, {Salim}, {Weinberg}, {Lu}, {Do}, {Dunn},
  {Matthews}, {Morris}, {Yelda}, {Becklin}, {Kremenek}, {Milosavljevic}, \&
  {Naiman}}]{2008ApJ...689.1044G}
{Ghez}, A.~M., {Salim}, S., {Weinberg}, N.~N., {et~al.} 2008, \apj, 689, 1044

\bibitem[{{Gwinn} {et~al.}(1997){Gwinn}, {Eubanks}, {Pyne}, {Birkinshaw}, \&
  {Matsakis}}]{1997ApJ...485...87G}
{Gwinn}, C.~R., {Eubanks}, T.~M., {Pyne}, T., {Birkinshaw}, M., \& {Matsakis},
  D.~N. 1997, \apj, 485, 87

\bibitem[{{Hinshaw} {et~al.}(2009){Hinshaw}, {Weiland}, {Hill}, {Odegard},
  {Larson}, {Bennett}, {Dunkley}, {Gold}, {Greason}, {Jarosik}, {Komatsu},
  {Nolta}, {Page}, {Spergel}, {Wollack}, {Halpern}, {Kogut}, {Limon}, {Meyer},
  {Tucker}, \& {Wright}}]{2009ApJS..180..225H}
{Hinshaw}, G., {Weiland}, J.~L., {Hill}, R.~S., {et~al.} 2009, \apjs, 180, 225

\bibitem[{Hobbs {et~al.}(2009)Hobbs, Holl, Lindegren, Raison, Klioner, \&
  Butkevich}]{IAU:6911396}
Hobbs, D., Holl, B., Lindegren, L., {et~al.} 2009, in Proceedings of the
  International Astronomical Union, Vol.~5, Relativity in Fundamental
  Astronomy: Dynamics, Reference Frames, and Data Analysis, 315--319

\bibitem[{{Hobson} {et~al.}(2006){Hobson}, {Efstathiou}, \&
  {Lasenby}}]{2006gere.book.....H}
{Hobson}, M.~P., {Efstathiou}, G.~P., \& {Lasenby}, A.~N. 2006, {General
  Relativity} (Cambridge University Press)

\bibitem[{{Hogg}(1999)}]{1999astro.ph..5116H}
{Hogg}, D.~W. 1999, ArXiv Astrophysics e-prints [\eprint{astro-ph/9905116}]

\bibitem[{{Holl} {et~al.}(2012){Holl}, {Lindegren}, \&
  {Hobbs}}]{2012A&A...543A..15H}
{Holl}, B., {Lindegren}, L., \& {Hobbs}, D. 2012, \aap, 543, A15

\bibitem[{{Hopp} \& {Materne}(1985)}]{1985A&AS...61...93H}
{Hopp}, U. \& {Materne}, J. 1985, \aaps, 61, 93

\bibitem[{{Kardashev}(1986)}]{1986AZh....63..845K}
{Kardashev}, N.~S. 1986, \azh, 63, 845

\bibitem[{{Klioner}(2003)}]{2003AJ....125.1580K}
{Klioner}, S.~A. 2003, \aj, 125, 1580

\bibitem[{{Klioner}(2004)}]{2004PhRvD..69l4001K}
{Klioner}, S.~A. 2004, \prd, 69, 124001

\bibitem[{{Kopeikin} \& {Makarov}(2006)}]{2006AJ....131.1471K}
{Kopeikin}, S.~M. \& {Makarov}, V.~V. 2006, \aj, 131, 1471

\bibitem[{{Kovalev} {et~al.}(2008){Kovalev}, {Lobanov}, {Pushkarev}, \&
  {Zensus}}]{2008A&A...483..759K}
{Kovalev}, Y.~Y., {Lobanov}, A.~P., {Pushkarev}, A.~B., \& {Zensus}, J.~A.
  2008, \aap, 483, 759

\bibitem[{{Kovalevsky}(2003)}]{2003A&A...404..743K}
{Kovalevsky}, J. 2003, \aap, 404, 743

\bibitem[{{Kristian} \& {Sachs}(1966)}]{1966ApJ...143..379K}
{Kristian}, J. \& {Sachs}, R.~K. 1966, \apj, 143, 379

\bibitem[{{Lewis} \& {Ibata}(1998)}]{1998ApJ...501..478L}
{Lewis}, G.~F. \& {Ibata}, R.~A. 1998, \apj, 501, 478

\bibitem[{{Lindegren} {et~al.}(2012){Lindegren}, {Lammers}, {Hobbs},
  {O'Mullane}, {Bastian}, \& {Hern{\'a}ndez}}]{2012A&A...538A..78L}
{Lindegren}, L., {Lammers}, U., {Hobbs}, D., {et~al.} 2012, \aap, 538, A78

\bibitem[{{Ma} {et~al.}(2009){Ma}, {Arias}, {Bianco}, {Boboltz}, {Bolotin},
  {Charlot}, {Engelhardt}, {Fey}, {Gaume}, {Gontier}, {Heinkelmann}, {Jacobs},
  {Kurdubov}, {Lambert}, {Malkin}, {Nothnagel}, {Petrov}, {Skurikhina},
  {Sokolova}, {Souchay}, {Sovers}, {Tesmer}, {Titov}, {Wang}, {Zharov},
  {Barache}, {Boeckmann}, {Collioud}, {Gipson}, {Gordon}, {Lytvyn},
  {MacMillan}, \& {Ojha}}]{2009ITN....35....1M}
{Ma}, C., {Arias}, E.~F., {Bianco}, G., {et~al.} 2009, IERS Technical Note, 35,
  1

\bibitem[{{Ma} {et~al.}(1998){Ma}, {Arias}, {Eubanks}, {Fey}, {Gontier},
  {Jacobs}, {Sovers}, {Archinal}, \& {Charlot}}]{1998AJ....116..516M}
{Ma}, C., {Arias}, E.~F., {Eubanks}, T.~M., {et~al.} 1998, \aj, 116, 516

\bibitem[{{MacMillan}(2005)}]{2005ASPC..340..477M}
{MacMillan}, D.~S. 2005, in Astronomical Society of the Pacific Conference
  Series, Vol. 340, Future Directions in High Resolution Astronomy, ed.
  J.~{Romney} \& M.~{Reid}, 477

\bibitem[{{Makarov} {et~al.}(2012){Makarov}, {Berghea}, {Boboltz}, {Dieck},
  {Dorland}, {Dudik}, {Fey}, {Gaume}, {Lei}, {Schmitt}, \&
  {Zacharias}}]{2012MmSAI..83..952M}
{Makarov}, V., {Berghea}, C., {Boboltz}, D., {et~al.} 2012, \memsai, 83, 952

\bibitem[{{Malkin}(2014)}]{2014MNRAS.445..845M}
{Malkin}, Z. 2014, \mnras, 445, 845

\bibitem[{{Mignard}(2011)}]{Mignard2011}
{Mignard}, F. 2011, in Proceedings of the "Journ\'ees 2011 Syst\'emes de
  r\'ef\'erence spatio-temporels", eds.: Schuh, H., B\"{o}hm S., Nilsson T. and
  Capitaine N.

\bibitem[{{Mignard} \& {Klioner}(2012)}]{2012A&A...547A..59M}
{Mignard}, F. \& {Klioner}, S. 2012, \aap, 547, A59

\bibitem[{{Nasonova} {et~al.}(2011){Nasonova}, {de Freitas Pacheco}, \&
  {Karachentsev}}]{2011A&A...532A.104N}
{Nasonova}, O.~G., {de Freitas Pacheco}, J.~A., \& {Karachentsev}, I.~D. 2011,
  \aap, 532, A104

\bibitem[{{Orosz} \& {Frey}(2013)}]{2013A&A...553A..13O}
{Orosz}, G. \& {Frey}, S. 2013, \aap, 553, A13

\bibitem[{{P{\^a}ris} {et~al.}(2014){P{\^a}ris}, {Petitjean}, {Aubourg},
  {Ross}, {Myers}, {Streblyanska}, {Bailey}, {Hall}, {Strauss}, {Anderson},
  {Bizyaev}, {Borde}, {Brinkmann}, {Bovy}, {Brandt}, {Brewington},
  {Brownstein}, {Cook}, {Ebelke}, {Fan}, {Filiz Ak}, {Finley}, {Font-Ribera},
  {Ge}, {Hamann}, {Ho}, {Jiang}, {Kinemuchi}, {Malanushenko}, {Malanushenko},
  {Marchante}, {McGreer}, {McMahon}, {Miralda-Escud{\'e}}, {Muna},
  {Noterdaeme}, {Oravetz}, {Palanque-Delabrouille}, {Pan}, {Perez-Fournon},
  {Pieri}, {Riffel}, {Schlegel}, {Schneider}, {Simmons}, {Viel}, {Weaver},
  {Wood-Vasey}, {Y{\`e}che}, \& {York}}]{2014A&A...563A..54P}
{P{\^a}ris}, I., {Petitjean}, P., {Aubourg}, {\'E}., {et~al.} 2014, \aap, 563,
  A54

\bibitem[{{Pasquato} {et~al.}(2011){Pasquato}, {Pourbaix}, \&
  {Jorissen}}]{2011A&A...532A..13P}
{Pasquato}, E., {Pourbaix}, D., \& {Jorissen}, A. 2011, \aap, 532, A13

\bibitem[{{Perryman} {et~al.}(2014){Perryman}, {Spergel}, \&
  {Lindegren}}]{2014ApJ...789..166P}
{Perryman}, M., {Spergel}, D.~N., \& {Lindegren}, L. 2014, \apj, 789, 166

\bibitem[{{Planck Collaboration} {et~al.}(2014){Planck Collaboration},
  {Aghanim}, {Armitage-Caplan}, {Arnaud}, {Ashdown}, {Atrio-Barandela},
  {Aumont}, {Baccigalupi}, {Banday}, {Barreiro}, {Bartlett}, {Benabed},
  {Benoit-L{\'e}vy}, {Bernard}, {Bersanelli}, {Bielewicz}, {Bobin}, {Bock},
  {Bond}, {Borrill}, {Bouchet}, {Bridges}, {Burigana}, {Butler}, {Cardoso},
  {Catalano}, {Challinor}, {Chamballu}, {Chiang}, {Chiang}, {Christensen},
  {Clements}, {Colombo}, {Couchot}, {Crill}, {Curto}, {Cuttaia}, {Danese},
  {Davies}, {Davis}, {de Bernardis}, {de Rosa}, {de Zotti}, {Delabrouille},
  {Diego}, {Donzelli}, {Dor{\'e}}, {Dupac}, {Efstathiou}, {En{\ss}lin},
  {Eriksen}, {Finelli}, {Forni}, {Frailis}, {Franceschi}, {Galeotta}, {Ganga},
  {Giard}, {Giardino}, {Gonz{\'a}lez-Nuevo}, {G{\'o}rski}, {Gratton},
  {Gregorio}, {Gruppuso}, {Hansen}, {Hanson}, {Harrison}, {Helou},
  {Hildebrandt}, {Hivon}, {Hobson}, {Holmes}, {Hovest}, {Huffenberger},
  {Jones}, {Juvela}, {Keih{\"a}nen}, {Keskitalo}, {Kisner}, {Knoche}, {Knox},
  {Kunz}, {Kurki-Suonio}, {L{\"a}hteenm{\"a}ki}, {Lamarre}, {Lasenby},
  {Laureijs}, {Lawrence}, {Leonardi}, {Lewis}, {Liguori}, {Lilje},
  {Linden-V{\o}rnle}, {L{\'o}pez-Caniego}, {Lubin}, {Mac{\'{\i}}as-P{\'e}rez},
  {Mandolesi}, {Maris}, {Marshall}, {Martin}, {Mart{\'{\i}}nez-Gonz{\'a}lez},
  {Masi}, {Massardi}, {Matarrese}, {Mazzotta}, {Meinhold}, {Melchiorri},
  {Mendes}, {Migliaccio}, {Mitra}, {Moneti}, {Montier}, {Morgante}, {Mortlock},
  {Moss}, {Munshi}, {Naselsky}, {Nati}, {Natoli}, {N{\o}rgaard-Nielsen},
  {Noviello}, {Novikov}, {Novikov}, {Osborne}, {Oxborrow}, {Pagano}, {Pajot},
  {Paoletti}, {Pasian}, {Patanchon}, {Perdereau}, {Perrotta}, {Piacentini},
  {Pierpaoli}, {Pietrobon}, {Plaszczynski}, {Pointecouteau}, {Polenta},
  {Ponthieu}, {Popa}, {Pratt}, {Pr{\'e}zeau}, {Puget}, {Rachen}, {Reach},
  {Reinecke}, {Ricciardi}, {Riller}, {Ristorcelli}, {Rocha}, {Rosset},
  {Rubi{\~n}o-Mart{\'{\i}}n}, {Rusholme}, {Santos}, {Savini}, {Scott},
  {Seiffert}, {Shellard}, {Spencer}, {Sunyaev}, {Sureau}, {Suur-Uski},
  {Sygnet}, {Tauber}, {Tavagnacco}, {Terenzi}, {Toffolatti}, {Tomasi},
  {Tristram}, {Tucci}, {T{\"u}rler}, {Valenziano}, {Valiviita}, {Van Tent},
  {Vielva}, {Villa}, {Vittorio}, {Wade}, {Wandelt}, {White}, {Yvon}, {Zacchei},
  {Zibin}, \& {Zonca}}]{2014A&A...571A..27P}
{Planck Collaboration}, {Aghanim}, N., {Armitage-Caplan}, C., {et~al.} 2014,
  \aap, 571, A27

\bibitem[{{Popovi{\'c}} {et~al.}(2012){Popovi{\'c}}, {Jovanovi{\'c}},
  {Stalevski}, {Anton}, {Andrei}, {Kova{\v c}evi{\'c}}, \&
  {Baes}}]{2012A&A...538A.107P}
{Popovi{\'c}}, L.~{\v C}., {Jovanovi{\'c}}, P., {Stalevski}, M., {et~al.} 2012,
  \aap, 538, A107

\bibitem[{{Porcas}(2009)}]{2009A&A...505L...1P}
{Porcas}, R.~W. 2009, \aap, 505, L1

\bibitem[{{Pyne} {et~al.}(1996){Pyne}, {Gwinn}, {Birkinshaw}, {Eubanks}, \&
  {Matsakis}}]{1996ApJ...465..566P}
{Pyne}, T., {Gwinn}, C.~R., {Birkinshaw}, M., {Eubanks}, T.~M., \& {Matsakis},
  D.~N. 1996, \apj, 465, 566

\bibitem[{{Quercellini} {et~al.}(2012){Quercellini}, {Amendola}, {Balbi},
  {Cabella}, \& {Quartin}}]{2012PhR...521...95Q}
{Quercellini}, C., {Amendola}, L., {Balbi}, A., {Cabella}, P., \& {Quartin}, M.
  2012, \physrep, 521, 95

\bibitem[{{Quercellini} {et~al.}(2009){Quercellini}, {Cabella}, {Amendola},
  {Quartin}, \& {Balbi}}]{2009PhRvD..80f3527Q}
{Quercellini}, C., {Cabella}, P., {Amendola}, L., {Quartin}, M., \& {Balbi}, A.
  2009, \prd, 80, 063527

\bibitem[{{Robin} {et~al.}(2012){Robin}, {Luri}, {Reyl{\'e}}, {Isasi}, {Grux},
  {Blanco-Cuaresma}, {Arenou}, {Babusiaux}, {Belcheva}, {Drimmel}, {Jordi},
  {Krone-Martins}, {Masana}, {Mauduit}, {Mignard}, {Mowlavi},
  {Rocca-Volmerange}, {Sartoretti}, {Slezak}, \&
  {Sozzetti}}]{2012A&A...543A.100R}
{Robin}, A.~C., {Luri}, X., {Reyl{\'e}}, C., {et~al.} 2012, \aap, 543, A100

\bibitem[{{Sazhin} {et~al.}(2011){Sazhin}, {Sazhina}, \&
  {Pshirkov}}]{2011ARep...55..954S}
{Sazhin}, M.~V., {Sazhina}, O.~S., \& {Pshirkov}, M.~S. 2011, Astronomy
  Reports, 55, 954

\bibitem[{{Secrest} {et~al.}(2015){Secrest}, {Dudik}, {Dorland}, {Zacharias},
  {Makarov}, {Fey}, {Frouard}, \& {Finch}}]{2015ApJS..221...12S}
{Secrest}, N.~J., {Dudik}, R.~P., {Dorland}, B.~N., {et~al.} 2015, \apjs, 221,
  12

\bibitem[{{Shen} {et~al.}(2011){Shen}, {Richards}, {Strauss}, {Hall},
  {Schneider}, {Snedden}, {Bizyaev}, {Brewington}, {Malanushenko},
  {Malanushenko}, {Oravetz}, {Pan}, \& {Simmons}}]{2011ApJS..194...45S}
{Shen}, Y., {Richards}, G.~T., {Strauss}, M.~A., {et~al.} 2011, \apjs, 194, 45

\bibitem[{{Smith} {et~al.}(1993){Smith}, {Nair}, {Leacock}, \&
  {Clements}}]{1993AJ....105..437S}
{Smith}, A.~G., {Nair}, A.~D., {Leacock}, R.~J., \& {Clements}, S.~D. 1993,
  \aj, 105, 437

\bibitem[{{Souchay} {et~al.}(2012){Souchay}, {Andrei}, {Barache}, {Bouquillon},
  {Suchet}, {Taris}, \& {Peralta}}]{2012A&A...537A..99S}
{Souchay}, J., {Andrei}, A.~H., {Barache}, C., {et~al.} 2012, \aap, 537, A99

\bibitem[{{Sovers} {et~al.}(1998){Sovers}, {Fanselow}, \&
  {Jacobs}}]{1998RvMP...70.1393S}
{Sovers}, O.~J., {Fanselow}, J.~L., \& {Jacobs}, C.~S. 1998, Reviews of Modern
  Physics, 70, 1393

\bibitem[{{Sparke} \& {Gallagher}(2000)}]{2000gaun.book.....S}
{Sparke}, L.~S. \& {Gallagher}, III, J.~S. 2000, {Galaxies in the universe : an
  introduction} (Cambridge University Press)

\bibitem[{{Taris} {et~al.}(2011){Taris}, {Souchay}, {Andrei}, {Bernard},
  {Salabert}, {Bouquillon}, {Anton}, {Lambert}, {Gontier}, \&
  {Barache}}]{2011A&A...526A..25T}
{Taris}, F., {Souchay}, J., {Andrei}, A.~H., {et~al.} 2011, \aap, 526, A25

\bibitem[{{Titov}(2010)}]{2010ivs..conf...60T}
{Titov}, O. 2010, in Sixth International VLBI Service for Geodesy and
  Astronomy. Proceedings from the 2010 General Meeting, ''VLBI2010: From Vision
  to Reality''. Held 7-13 February, 2010 in Hobart, Tasmania, Australia. Edited
  by D. Behrend and K.D. Baver. NASA/CP 2010-215864., p.60-64, ed.
  R.~{Navarro}, S.~{Rogstad}, C.~E. {Goodhart}, E.~{Sigman}, M.~{Soriano},
  D.~{Wang}, L.~A. {White}, \& C.~S. {Jacobs}, 60--64

\bibitem[{{Titov} {et~al.}(2011){Titov}, {Lambert}, \&
  {Gontier}}]{2011A&A...529A..91T}
{Titov}, O., {Lambert}, S.~B., \& {Gontier}, A.-M. 2011, \aap, 529, A91

\bibitem[{Van~den Bergh {et~al.}(2000)Van~den Bergh, King, Lin, Maran, Pringle,
  \& Ward}]{VandenBergh:991671}
Van~den Bergh, S., King, A., Lin, D., {et~al.} 2000, {Galaxies of the Local
  Group} (Cambridge: Cambridge Univ. Press)

\bibitem[{{V{\'e}ron-Cetty} \& {V{\'e}ron}(2010)}]{2010A&A...518A..10V}
{V{\'e}ron-Cetty}, M.-P. \& {V{\'e}ron}, P. 2010, \aap, 518, A10

\bibitem[{{Weinberg}(2008)}]{2008cosm.book.....W}
{Weinberg}, S. 2008, {Cosmology} (Oxford University Press)

\bibitem[{{Zacharias} \& {Zacharias}(2014)}]{2014AJ....147...95Z}
{Zacharias}, N. \& {Zacharias}, M.~I. 2014, \aj, 147, 95

\end{thebibliography}
\end{document}